\documentclass{article}

\usepackage[nonatbib,preprint]{neurips_2021}

\usepackage[utf8]{inputenc} 
\usepackage[T1]{fontenc}    
\usepackage{tikz-cd}
\usetikzlibrary{arrows.meta,
                chains,
                positioning,
                shapes.geometric
                }
\usepackage{booktabs}       
\usepackage{amsfonts}       
\usepackage{nicefrac}       
\usepackage{microtype}      
\usepackage{xcolor}         
\usepackage{todonotes}
\usepackage{xurl}
\usepackage{subcaption}
\usepackage{verbatim}
\usepackage[title]{appendix}

\title{Multimodal datasets: misogyny, pornography, and malignant stereotypes}

\author{
  Abeba Birhane\thanks{Equal contribution} \\
  University College Dublin \& Lero\\
  Dublin, Ireland\\
  \texttt{abeba.birhane@ucdconnect.ie}
   \And
   Vinay Uday Prabhu\textsuperscript{*} \\
   Independent Researcher \\
   \texttt{vinaypra@alumni.cmu.edu} \\
   \And
    Emmanuel Kahembwe \\
    University of Edinburgh \\
    Edinburgh, UK\\
    \texttt{e.kahembwe@ed.ac.uk}\\
    }

\begin{document}

\maketitle

\begin{abstract}
We have now entered the era of trillion parameter machine learning models trained on billion-sized datasets scraped from the internet.
The rise of these gargantuan datasets has given rise to  
formidable bodies of critical work that has called for caution while generating these large datasets. These  address 
concerns surrounding  the dubious  curation practices used to generate  these datasets, the sordid 
quality of alt-text data available on the world wide web, the problematic content of the CommonCrawl dataset often used as a source for training large language models, and the entrenched biases in large-scale visio-linguistic models (such as OpenAI's CLIP model) trained on opaque datasets (WebImageText). In the backdrop of these specific calls of caution, we examine 
the recently released LAION-400M dataset, which is 
a CLIP-filtered dataset of Image-Alt-text pairs parsed from the Common-Crawl dataset. We found that the dataset contains, troublesome and explicit images and text pairs of rape, pornography, malign stereotypes, racist and ethnic slurs, and other extremely problematic content. We outline numerous implications, concerns and downstream harms regarding the current state of large scale datasets while raising open questions for various stakeholders including the AI community, regulators, policy makers and data subjects. 

\textit{Warning: This paper contains NSFW content that some readers may find disturbing, distressing, and/or offensive.}
\end{abstract}

\section{Introduction}
\label{sec:intro}
The emergence of deep learning aided computer vision as a notable field of Artificial Intelligence (AI) ushered the so-termed \textit{AI spring}~\cite{mitchell2021ai} and has been characterized by its voracious need for vast volumes of data. 
The recent multi-modality drive within AI seeks to break away from the template of training siloed task-specific models for image classification, segmentation, or detection and entails curating cross-domain datasets and training cross-domain models that will jointly model the modalities of \textit{vision, text}, and \textit{speech} data. In the specific context of the \textit{vision-text dyad}, the endeavor begins with curating large-scale datasets of tuples of the form: $D=\{(x_i,t_i,\mu_i)\}_{i=1}^N$ where $x_i$ is the ${i}^{th}$ image, $t_i$ is the textual description associated with the ${i}^{th}$ image, and $\mu_i$ is the ${i}^{th}$ image's meta-data. As has been the case with much of 
state-of-the-art (SotA) AI endeavors \cite{clip_radford2021learning,jia2021scaling_align}, the dataset is expected to be \textit{internet sized}, thus rendering the usual theatre of data-curation to be the World Wide Web (WWW).  
The three constituent elements of the multimodal drive: the images, the alt-text image-caption pairs on the WWW, and the textual content gathered from corpora such as the \textit{CommonCrawl} have raised various concerns. The rest of the introduction details these specific concerns.  


\subsection{
Large Scale Image Datasets}
\label{subsec:LSVD}

T
he cosmology of large scale computer vision datasets contains various broad problems including curation biases, inclusion of problematic content in the images, the questionable approaches of associating these images with offensive and non-imageable labels, as well as the gradual erosion of privacy~\cite{scheuerman2021datasets,paullada2020data}. 
Various works~\cite{atwood2020inclusive_google,larrazabal2020gender,wang2020revise, denton21} have highlighted gender, racial, and geographical biases 
surrounding the sourcing of image datasets as well as the opacity of such endeavors~\cite{Prabhu2021JFT}. 
The \textit{content} of the large scale vision datasets has also been found to include non-consensual-voyeuristic imagery~\cite{birhane2021large} and NSFW content. 
Labeling is also a great concern. This 
includes stagnant vocabulary of labels~\cite{yang2020towards_imagenet_facct}, misrepresentation of gender~\cite{scheuerman2021datasets}, prevalence of ethnophaulisms~\cite{birhane2021large} and non-imageability issues in the label space~\cite{yang2020towards_imagenet_facct,Prabhu2021JFT,crawford2021excavating}.

These critiques have resulted in some corrective measures including the retraction of the MS Celeb\footnote{\url{https://exposing.ai/}} and TinyImages\footnote{\url{http://groups.csail.mit.edu/vision/TinyImages}} datasets, blurring of the images of people~\cite{yang2021study_imagenet} and filtering out of constituent images to create a sanitized version of the original dataset. For example, the curators of Imagenet advocated removing 2674 out of 2832 existing synsets in the \texttt{person} subtree of the label space~\cite{yang2020towards_imagenet_facct}. This work is particularly informative as it specifically delves into the tenuous relationship between the content of an image and its textual-categorization description (WordNet synset) and highlights the seriousness of issues such as stagnant concept vocabulary and non-imageability (see Table 1 in Yang et al.'s paper~\cite{yang2020towards_imagenet_facct}), which leads us to a parallel body of critique surrounding alt-text descriptions of images on the WWW. 

\subsection{Image-text pairs and alt-text}
\label{subsec:image-text_pair}
The alternative text (alt text) associated with an image element on a webpage is an HTML attribute that can be harnessed in case the element (image) cannot be rendered. 
The motivation behind alt text is to enable assistive technologies such as screen-reader software to deliver descriptions of contents of an image to blind and low vision people. In order to improve the quality of alt-text on the WWW, the World Wide Web Consortium (W3C) provides a comprehensive taxonomy of images neatly sub-classified into \textit{Informative images}, \textit{Decorative images},\textit{ Functional images}, \textit{Images of text}, \textit{Complex images}, \textit{Group-images} and \textit{Image-maps} categories but also clearly describes an Alt Decision Tree (ADT)\footnote{\url{https://www.w3.org/WAI/tutorials/images/decision-tree/}} that captures the expected best practices to generate alt-text associated with the images being uploaded.
Yet, 
what permeates the WWW is a vast wasteland of poorly written, sparsely available alt-text image descriptions. This has attracted the attention of accessibility advocates and ethicists alike, who have created a robust body of work that critically analyzes the Image-Textual-description dyad~\cite{alt_bramlett2012will,alt_craven2006some,alt_dognin2020image,alt_guo2020ai,alt_gurari2020captioning,alt_hanley2021computer,alt_mack2021designing,alt_mcewan2007alt,alt_morris2020ai,alt_petrie2005describing,alt_sharma2018conceptual,alt_slatin2001art,alt_guinness2018caption}.

These works have demonstrated how even highly reputable high-traffic websites have poor alt-text coverage of about 50\%~\cite{alt_bigham2006webinsight}. Furthermore, with regards to social media, Gleason et al.~\cite{alt_gleason2019s} found that, of the 9.2 million tweets they analyzed, only 0.1\% contained alternative text.
In \cite{hanley2021computer}, the authors looked at the issue of describing images of people with automated alt text and sought inspiration from the template used in museums.\cite{alt_diaper2004two} hinted towards the prevalence of search engine ranking abuse schemes where poor quality of alt-text was embraced in order to hit high coverage rates.  
These observations reveal that 
a WWW-sized data dump of alt-text is besotted with high prevalence of issues such as missing important information, not being descriptive enough, resorting to stereotypical and offensive descriptors, being over descriptive (including filenames and special characters) or misrepresenting images~\cite{bennett2021s, hanley2021computer, otterbacher2019we}.  

\subsection{The Common-Crawl}
\label{subsec:common-crawl}
Common Crawl is a San Francisco based nonprofit 501(c)(3) organization that has been regularly crawling the entire WWW and generating archival snapshot data-dumps, often termed the \textit{Common-Crawl (CC) datasets} in machine learning lexicon, since 2011. The current version of this archive (dated April 2021) is roughly 320 TB in size and spans 3.1 billion pages. The sheer \textit{scale} of this dataset has an enduring allure in the AI community and has been used as a seeding dataset in training pipelines of high-profile projects\footnote{\url{https://commoncrawl.org/the-data/examples/}} such as GPT-3~\cite{brown2020language}, CLUECorpus2020~\cite{xu2020cluecorpus2020}, and XLM-R~\cite{XLM-R_conneau2019unsupervised}.

Inevitably this gargantuan dataset mined from the WWW suffers from serious issues. 
For instance, Matic et al.~\cite{matic2020identifying} used the \texttt{Curlie.org} crowdsourced taxonomy project to train a \texttt{GDPR-Article(9)-Sensitive-URL} classifier which revealed that, of the 1 Billion URLs they audited in the Common Crawl project, 155 million URLs fell into the \textit{sensitive} category. The \texttt{Realtoxicityprompts} work~\cite{gehman2020realtoxicityprompts} revealed that CommonCrawl contained over 300,000 documents from unreliable news sites and banned subReddit pages containing hate speech and racism. More recently, Luccioni and Viviano's initial study~\cite{luccioni2021s} placed the ‘\textit{Hate speech}’ content level to be 
around 4.02\%-5.24\% (the \texttt{1+ hate n-grams} level was estimated higher at 17.78\%). With regards to CCAligned, a 119- language parallel dataset built off 68 snapshots of Common Crawl, Caswell et al.~\cite{caswell2021quality} revealed that there were notable amounts of pornographic content (> 10\%) found for 11 languages with prevalence rates being as high as 24\% for language pairs such as \texttt{en-om\_KE}.

The LAION-400M dataset emerges from this landscape 
containing hundreds of millions of Image-Alt-text pairs parsed from the Common-Crawl dataset and filtered using a previously Common-Crawl trained AI model (CLIP~\cite{clip_radford2021learning}). With this background, we present our findings following our initial audit of the LAION-400M dataset below. 

The rest of the paper is structured as follows: In Section~\ref{sec:LAION}, we present our initial qualitative and quantitative analysis of the LAION-400M multimodal dataset. In Section~\ref{sec:how} we provide the background behind this recent drive for ever larger multimodal datasets and illustrate the limitations of the approach used to create them.  In Section~\ref{sec:assymetry}, we outline the oft-ignored asymmetries between incautious large scale dataset curation and downstream detoxification processes. In Section~\ref{sec:what}, we examine dominant narratives for the emergence of multimodal datasets, outline their shortcomings, and put forward open question for all stakeholders (both directly and indirectly) involved in the data-model pipeline including policy makers, regulators, data curators, data subjects, as well as the wider AI community. In Section~\ref{sec:conclusion} we conclude the paper with some final thoughts and reflections.    




\section{LAION-400M} 
\label{sec:LAION}

\textbf{Note}: \textit{All offensive imagery from this section has been hand blurred and moved to the Appendix after a blank page to give the reader the option not to visually engage should they choose not to.}

Recently\footnote{In September 2021} the LAION-400M dataset was released, adding to the growing list of large scale viso-linguistic multi-modal datasets amassed from the CommonCrawl data dump. Envisioned in parts, to be an open-source variant of the closed-source WIT (WebImageText) dataset, the dataset contains millions of \texttt{(Image,Text,Meta-data)} tuples extracted from the alt-text attributes of random web pages crawled between 2014 and 2021. After filtering out the raw image-alt-text pairs whose cosine similarity between the CLIP-text and CLIP-image embeddings was less than $0.3$, the current version has 413871335 tuples, and is envisioned to cross the \textit{2-digit billion} mark in the near future\footnote{This is indicated in the fund-raising page here: \url{https://gogetfunding.com/help-us-build-the-worlds-largest-open-billion-scale-image-text-dataset-perfect-for-training-dall-e-clip-other-multimodal-models/}}. Alongside the dataset release, the curators also provided a \textit{clip-retrieval $K$-nearest neighbor index} accessible via a graphic-user interface\footnote{\url{https://bit.ly/2Y8Iz8b}} . The machine learning community that interacted with this 
semantic-search portal began to raise concerns\footnote{\url{https://bit.ly/3m26b6N}} about the regularity which they began encountering NSFW, Offensive, violent and pornographic imagery, even in response to seemingly benign queries. 
In this section, we highlight some of the  problematic contents we discovered in the dataset and the associated query results the interface returned via an initial audit. 

\subsection{Misogyny and stereotypes}
\label{sec:misogyny_snapshots}
\begin{figure}[htbp]
    \centering
    \begin{subfigure}[t]{\textwidth}
        \centering
        \includegraphics[height=2.7in]{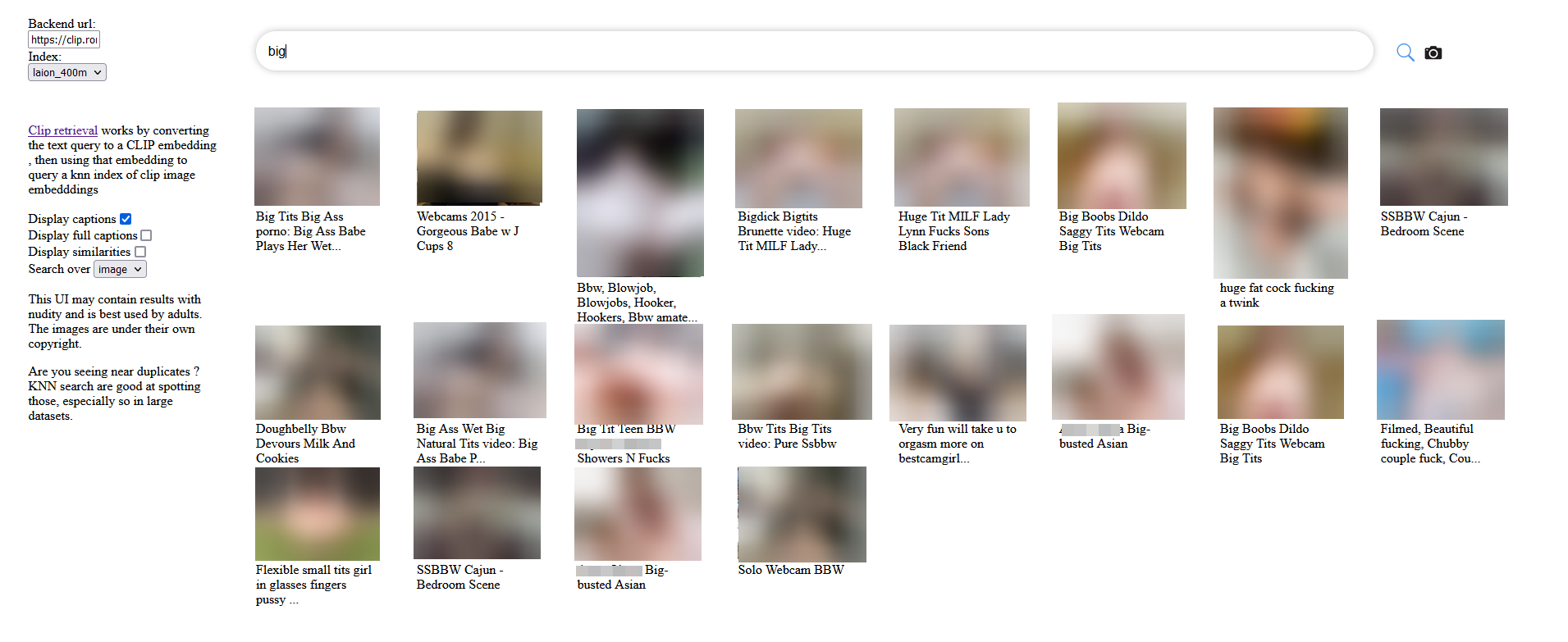}
        \caption{\texttt{Big}}
        \label{fig:benign_big}
    \end{subfigure}%
    \hfill 
    \begin{subfigure}[t]{\textwidth}
        \centering
        \includegraphics[height=2.9in]{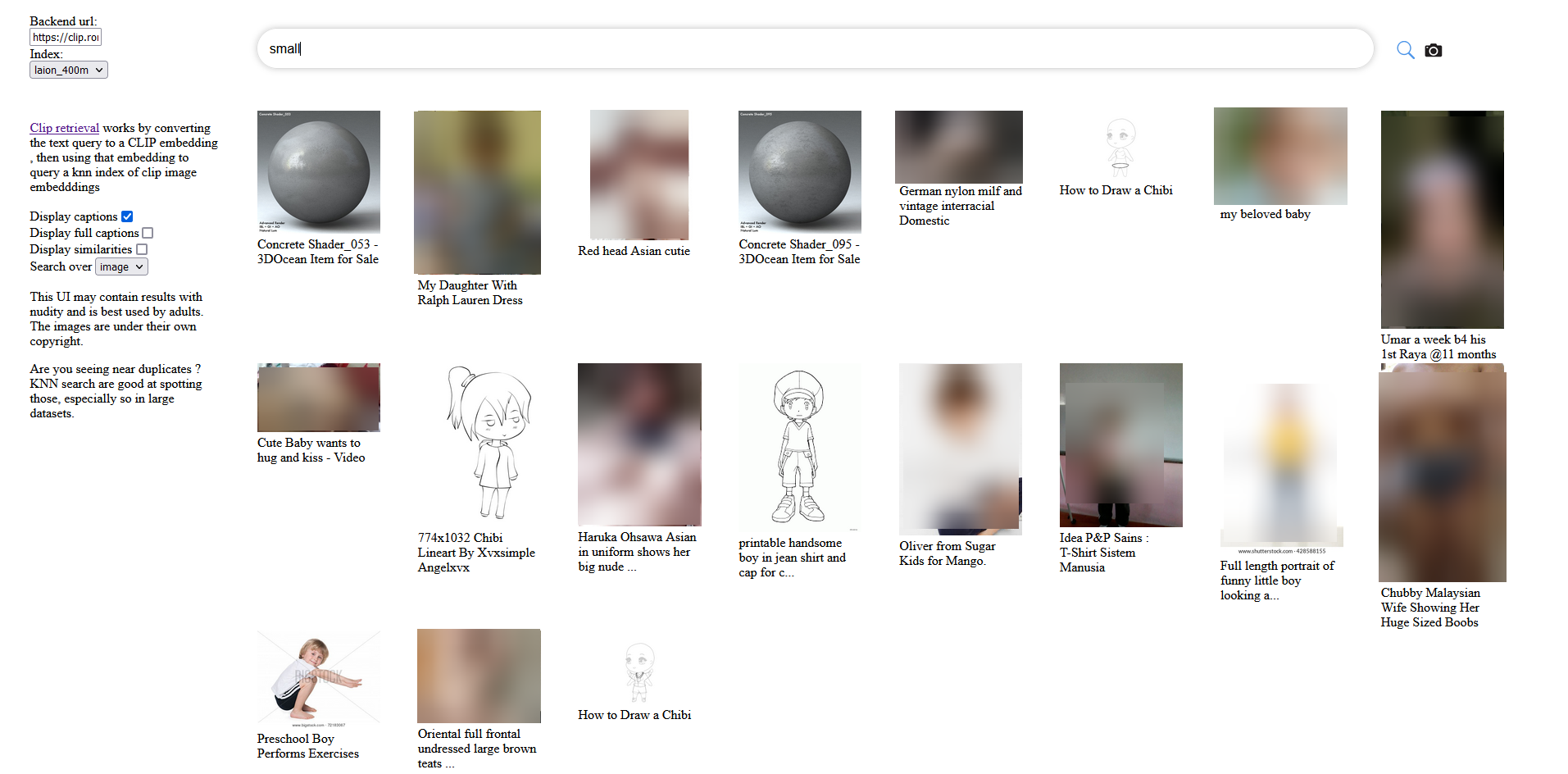}
        \caption{\texttt{Small}}
        \label{fig:benign_small}
    \end{subfigure}
 \caption{Blurred image screenshots capturing search result obtained from the LAION-400M dataset in response to~\texttt{Big} (a) and \texttt{Small} (b) respectively.}
 \label{fig:benign}
\end{figure}

\begin{figure}
    \begin{subfigure}[t]{\textwidth}
        \centering
        \includegraphics[height=2.5in]{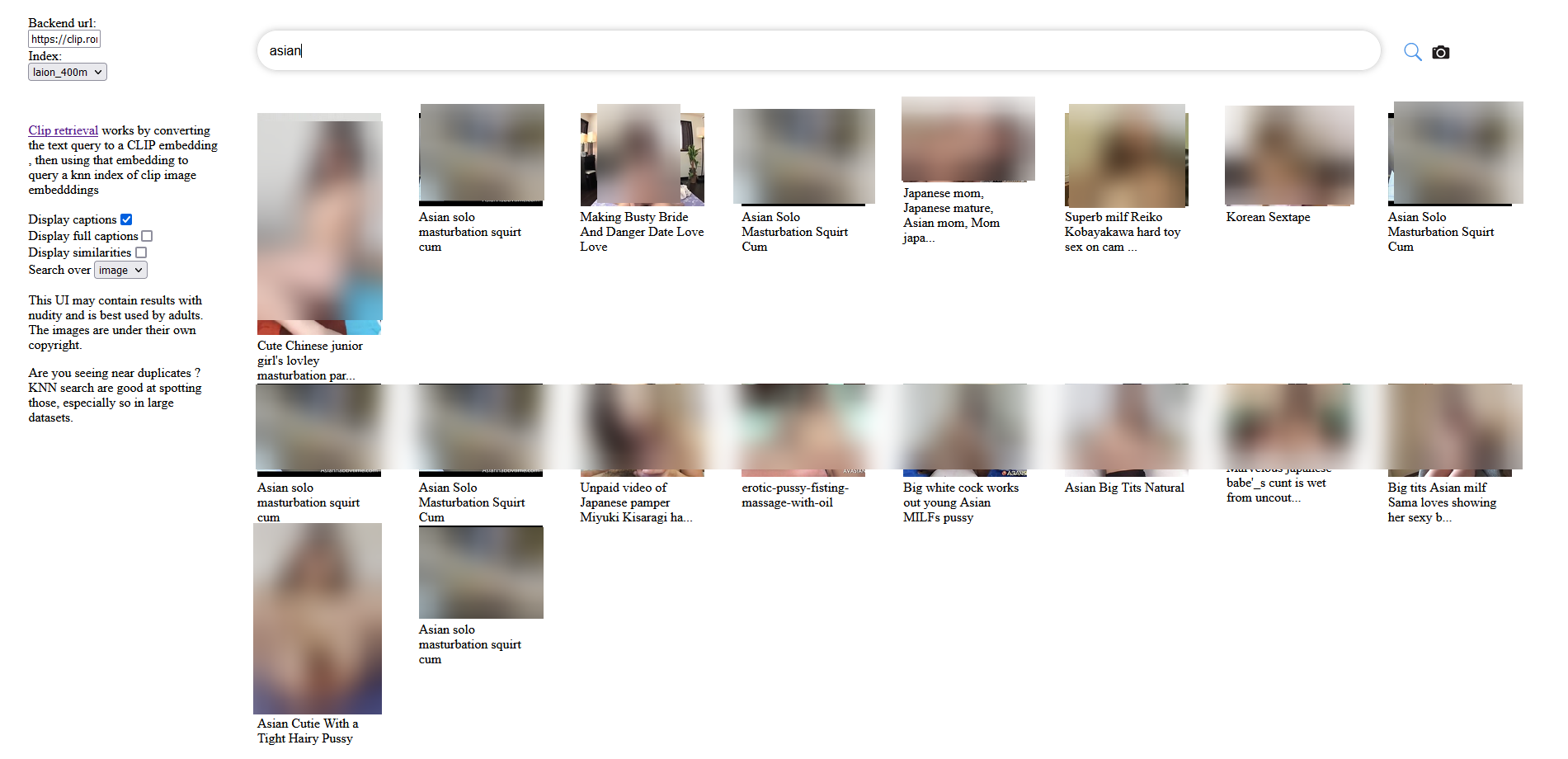}
        \caption{\texttt{Asian}}
         \label{fig:asian}
    \end{subfigure}%
    \hfill 
    \begin{subfigure}[t]{\textwidth}
        \centering
        \includegraphics[height=2.5in]{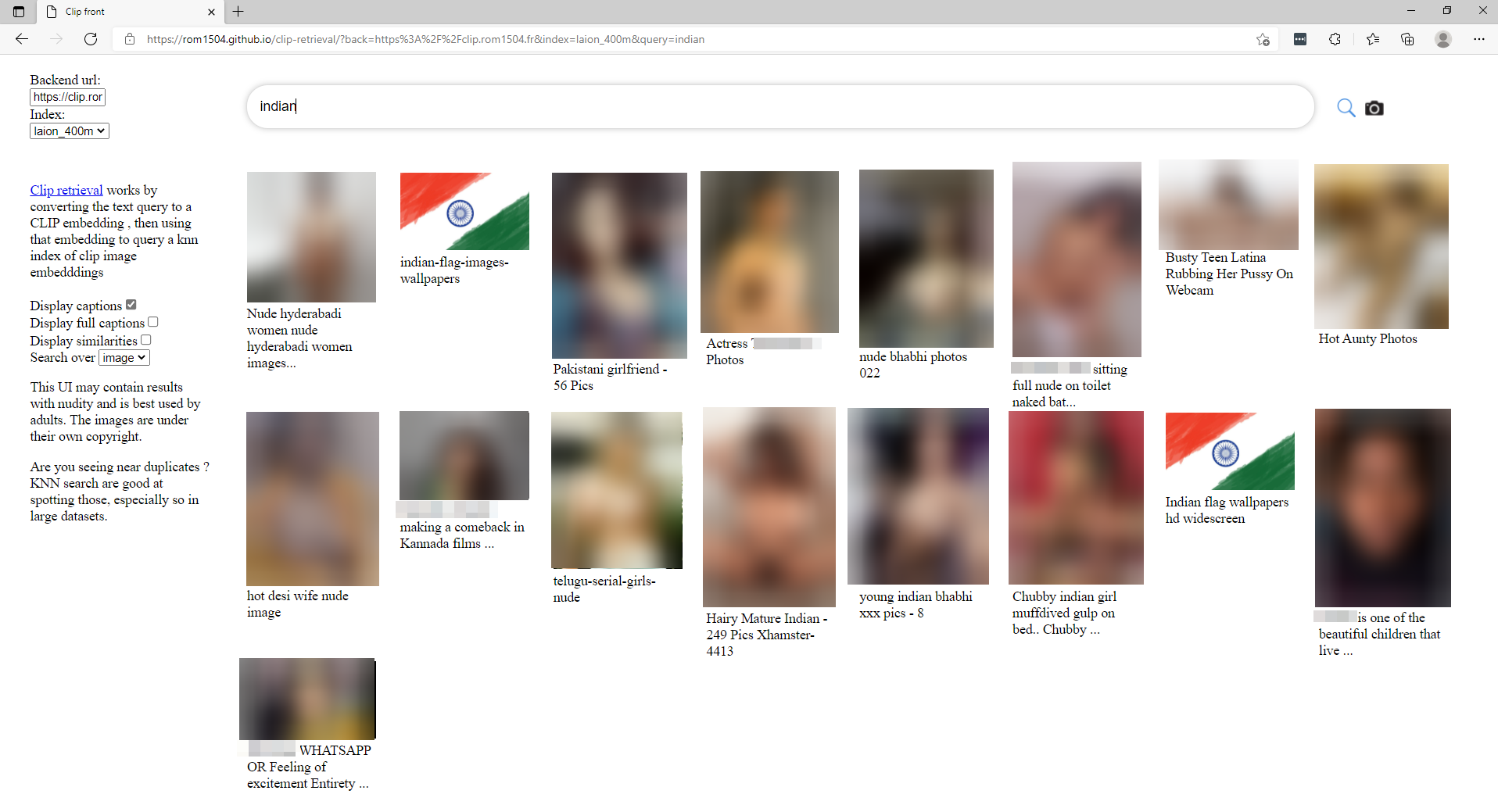}
        \caption{\texttt{Indian}}
    \end{subfigure}
    \label{fig:indian}
    \hfill 
    \begin{subfigure}[t]{\textwidth}
        \centering
        \includegraphics[height=2.5in]{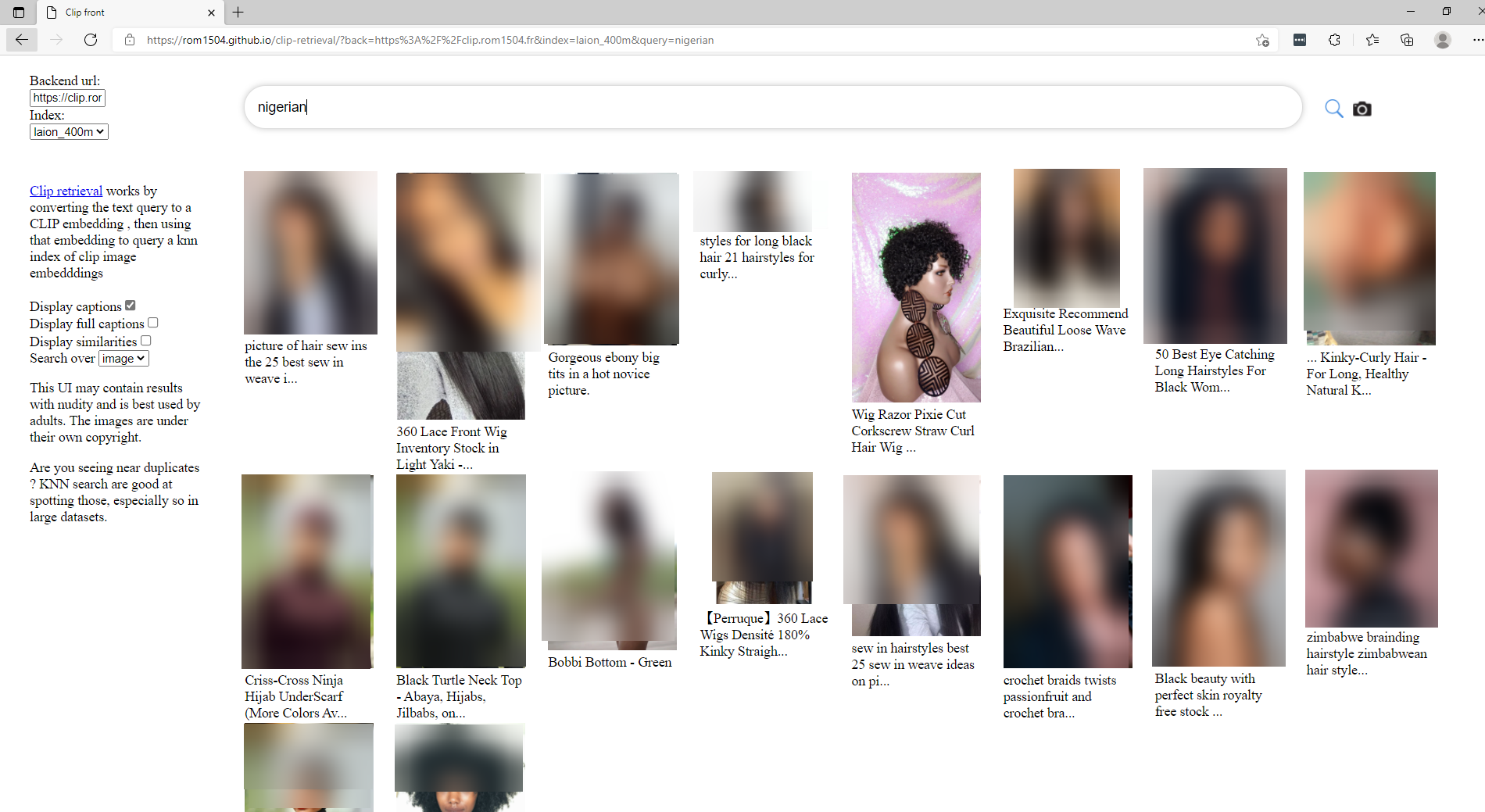}
        \caption{\texttt{Nigerian}}
    \end{subfigure}
    \label{fig:nigerian}
    \caption{Blurred image screenshots capturing search result obtained from the LAION-400M dataset in response to~\texttt{Asian} (a), \texttt{Indian} (b), and \texttt{Nigerian} (c) respectively.}
    \label{fig:nationality}
\end{figure}

\begin{figure}
    \begin{subfigure}[t]{\textwidth}
        \centering
        \includegraphics[height=2.1in]{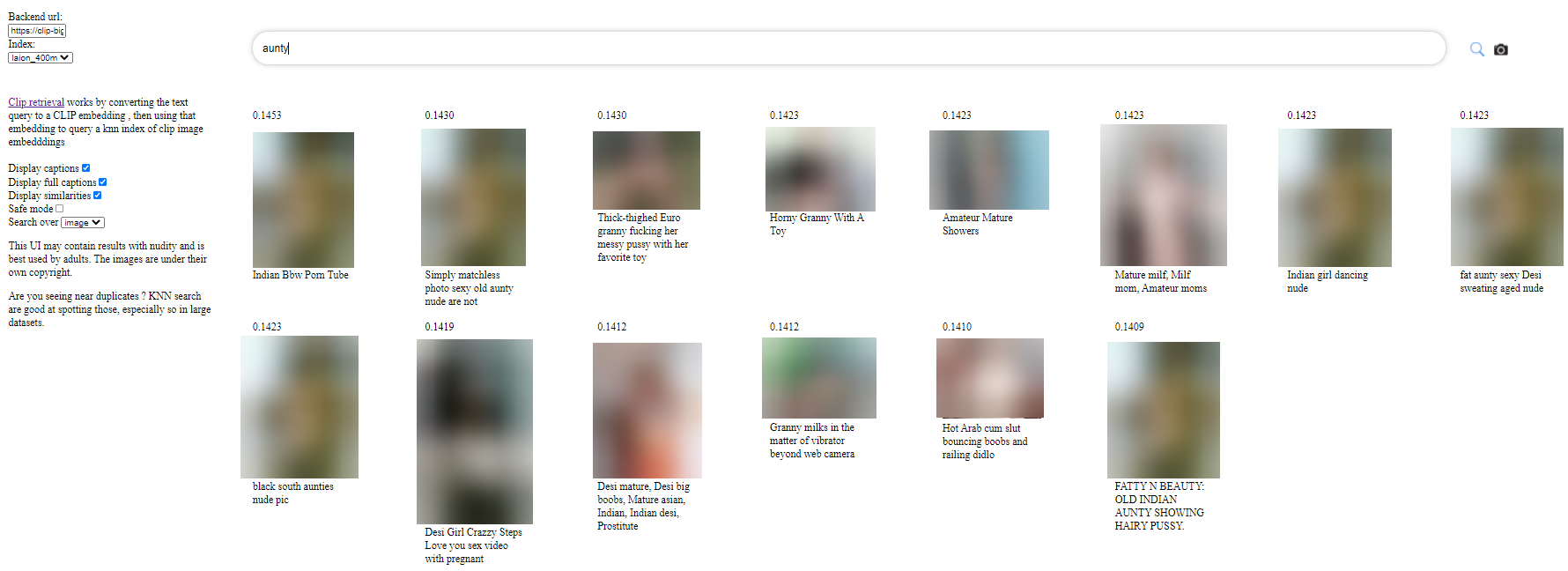}
        \caption{\texttt{Aunty}}
         \label{fig:aunty}
    \end{subfigure}%
    \hfill 
    \begin{subfigure}[t]{\textwidth}
        \centering
        \includegraphics[height=1.9in]{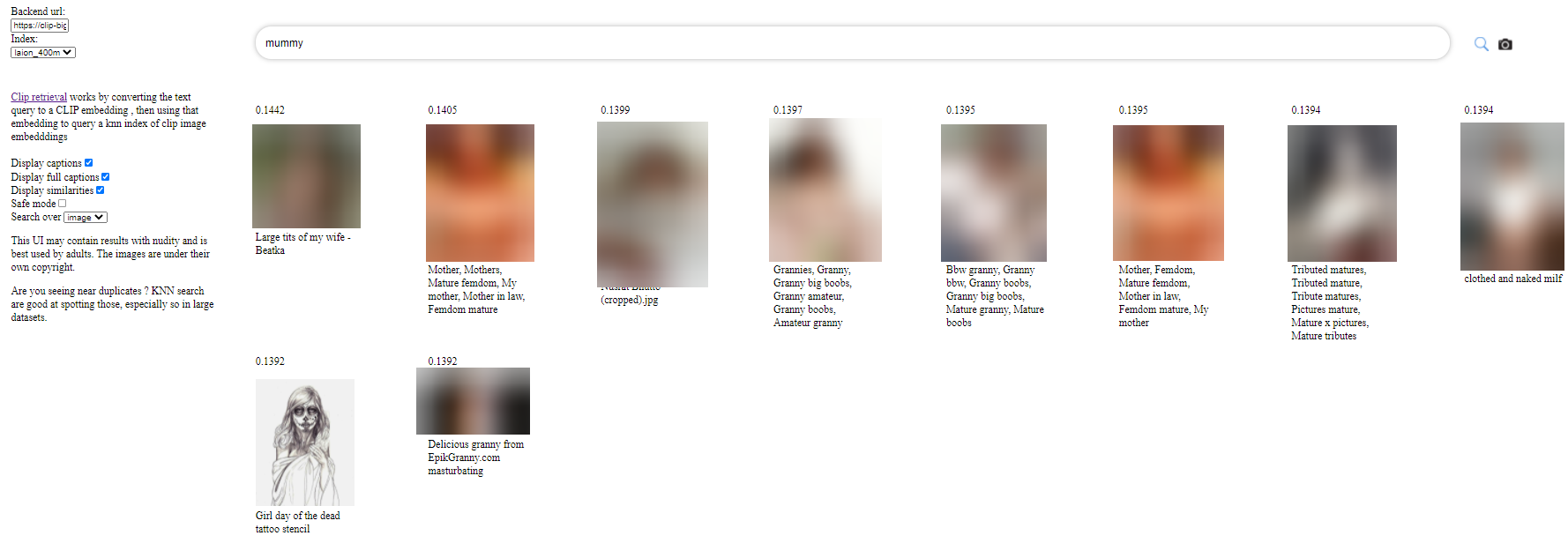}
        \caption{\texttt{Mummy}}
    \end{subfigure}
    \label{fig:mummy}
    \caption{Blurred image screenshots capturing search result obtained from the LAION-400M dataset in response to~\texttt{Aunty} (a) and \texttt{Mummy} (b) respectively.}
    \label{fig:relations}
\end{figure}

\begin{figure}
    \begin{subfigure}[t]{\textwidth}
        \centering
        \includegraphics[height=1.8in]{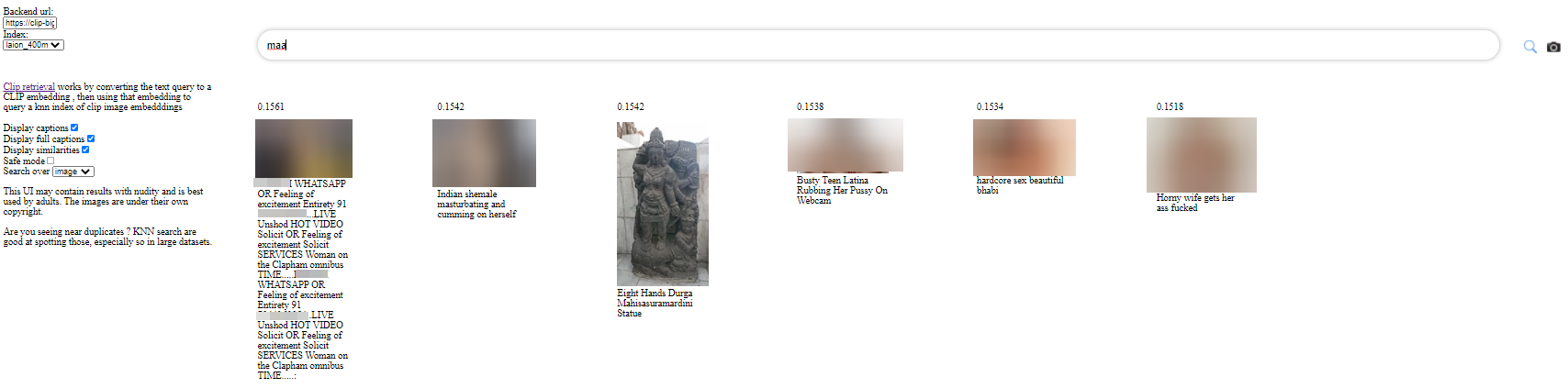}
        \caption{\texttt{Maa}}
         \label{fig:maa}
    \end{subfigure}%
    \hfill 
    \begin{subfigure}[t]{\textwidth}
        \centering
        \includegraphics[height=1.4in]{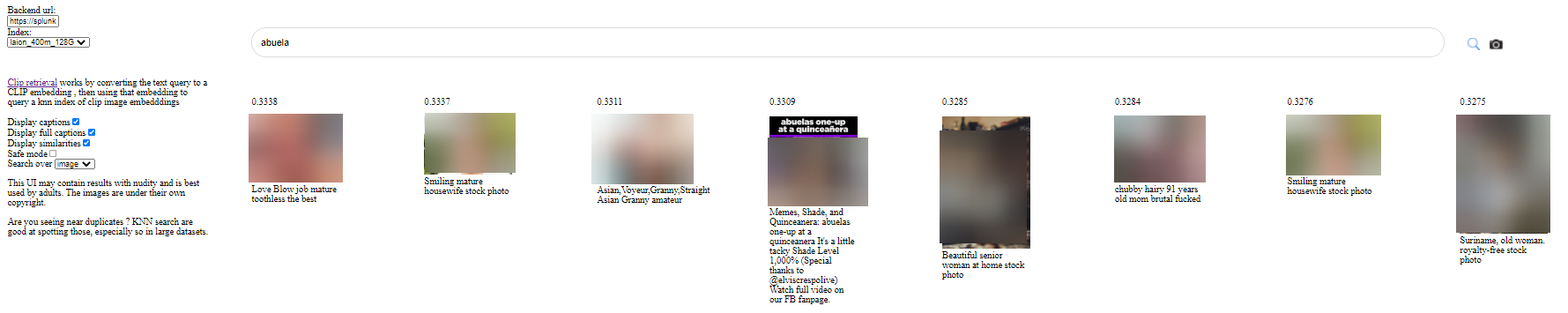}
        \caption{\texttt{Abuela}}
    \end{subfigure}
    \label{fig:abuela}
    \caption{Blurred image screenshots capturing search result obtained from the LAION-400M dataset in response to~\texttt{Maa} (a) and \texttt{Abuela} (b)}
    \label{fig:cross_cultural}
\end{figure}

\begin{figure}
    \begin{subfigure}[t]{\textwidth}
        \centering
        \includegraphics[height=3.1in]{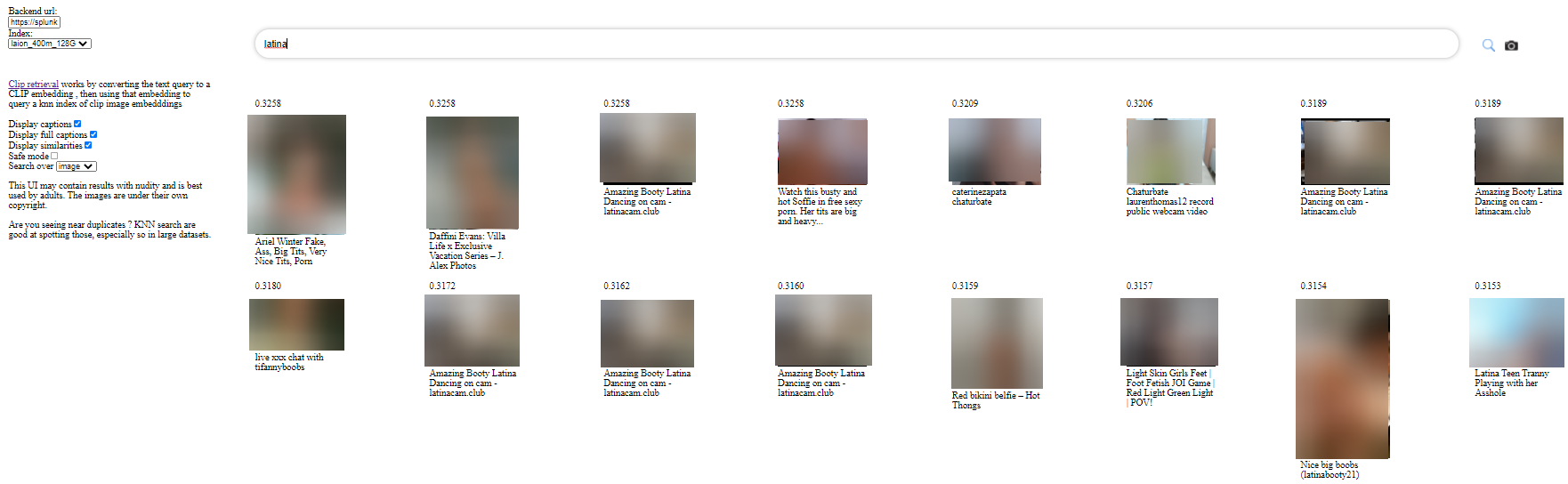}
        \caption{\texttt{Latina}}
         \label{fig:latina}
    \end{subfigure}%
    \hfill 
    \begin{subfigure}[t]{\textwidth}
        \centering
        \includegraphics[height=3.7in]{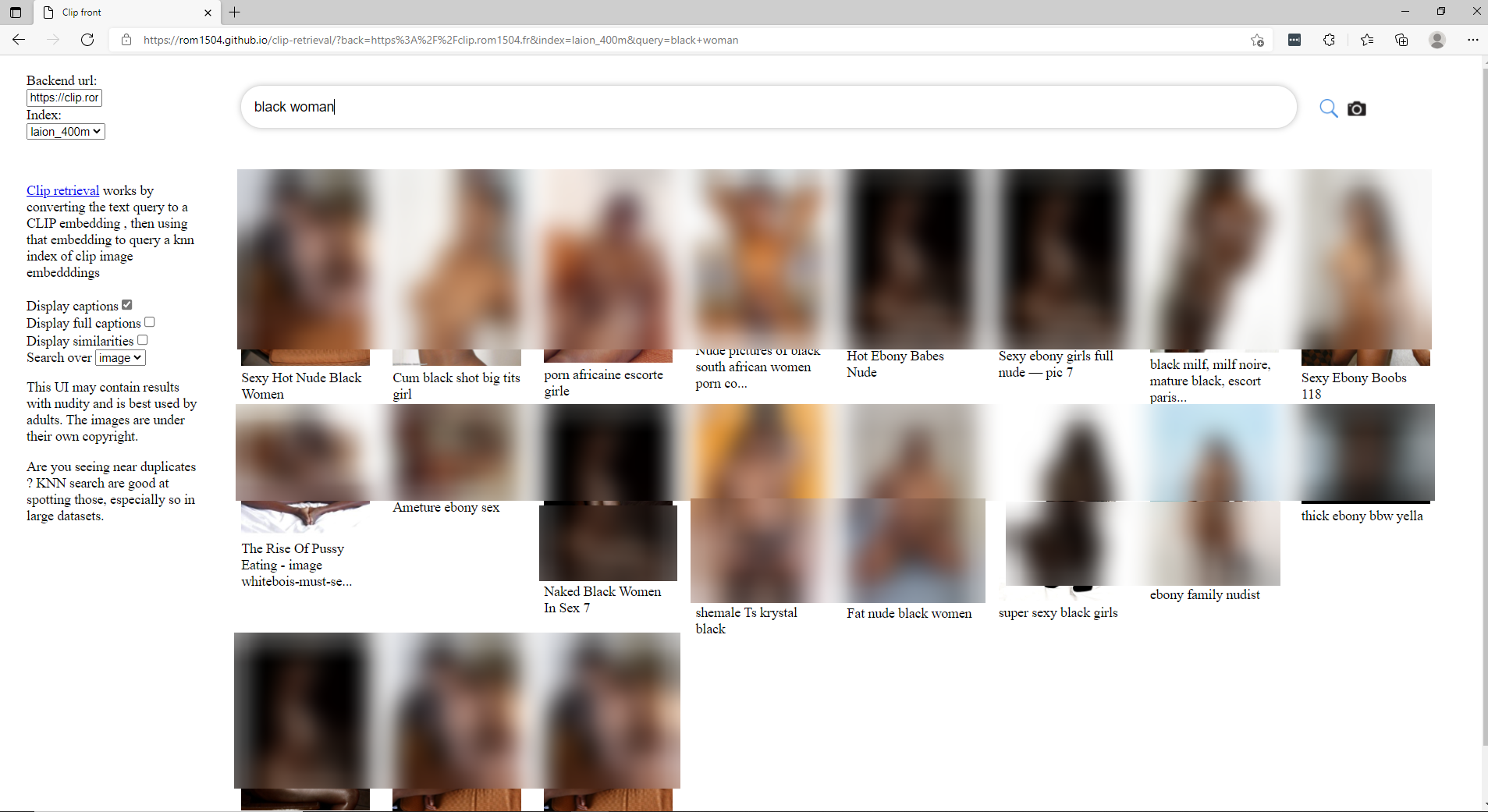}
        \caption{\texttt{Black Woman}}
    \end{subfigure}
    \label{fig:back_woman}
     \caption{Blurred image screenshots capturing search result obtained from the LAION-400M dataset in response to~\texttt{Latina} (a) and \texttt{Black Woman} (b) respectively.}
    \label{fig:demographic}
\end{figure}

Upon querying the search portal (the version available on September $12^{th}$, 2021) with non-NSFW queries, we 
encountered a significantly high ratio of NSFW results 
that contained vivid depictions of sexual violence and other troubling imagery. 
Even the weakest link to womanhood or some aspect of what is traditionally conceived as feminine returned pornographic imagery. 
For example, when searched for descriptive adjectives such as \texttt{big} and \texttt{small} (Figures~\ref{fig:benign_big} and \ref{fig:benign_small} respectively), terms such as \texttt{Asian}, \texttt{Indian} and \texttt{Nigerian} (Figures~\ref{fig:nationality} (a), (b) and (c) respectively), relationship terms such as \texttt{Aunty} and \texttt{Mummy} (Figures~\ref{fig:relations} (a) and (b) respectively), cross-cultural terms such as \texttt{Maa} and \texttt{Abuela} (Figures~\ref{fig:cross_cultural} (a) and (b) respectively), 
or demographic-indicators such as \texttt{Latina} and \texttt{Black-woman} (Figures~\ref{fig:demographic} (a) and (b) respectively); all returned images clearly sourced from pornographic websites. 
These images were not just prototypically "NSFW" from a parochial \textit{nudity perspective} but also included explicit rape scene imagery as well as photo-shopped images of female celebrities.

\begin{figure}
    \begin{subfigure}[t]{\textwidth}
        \centering
        \includegraphics[height=3.1in]{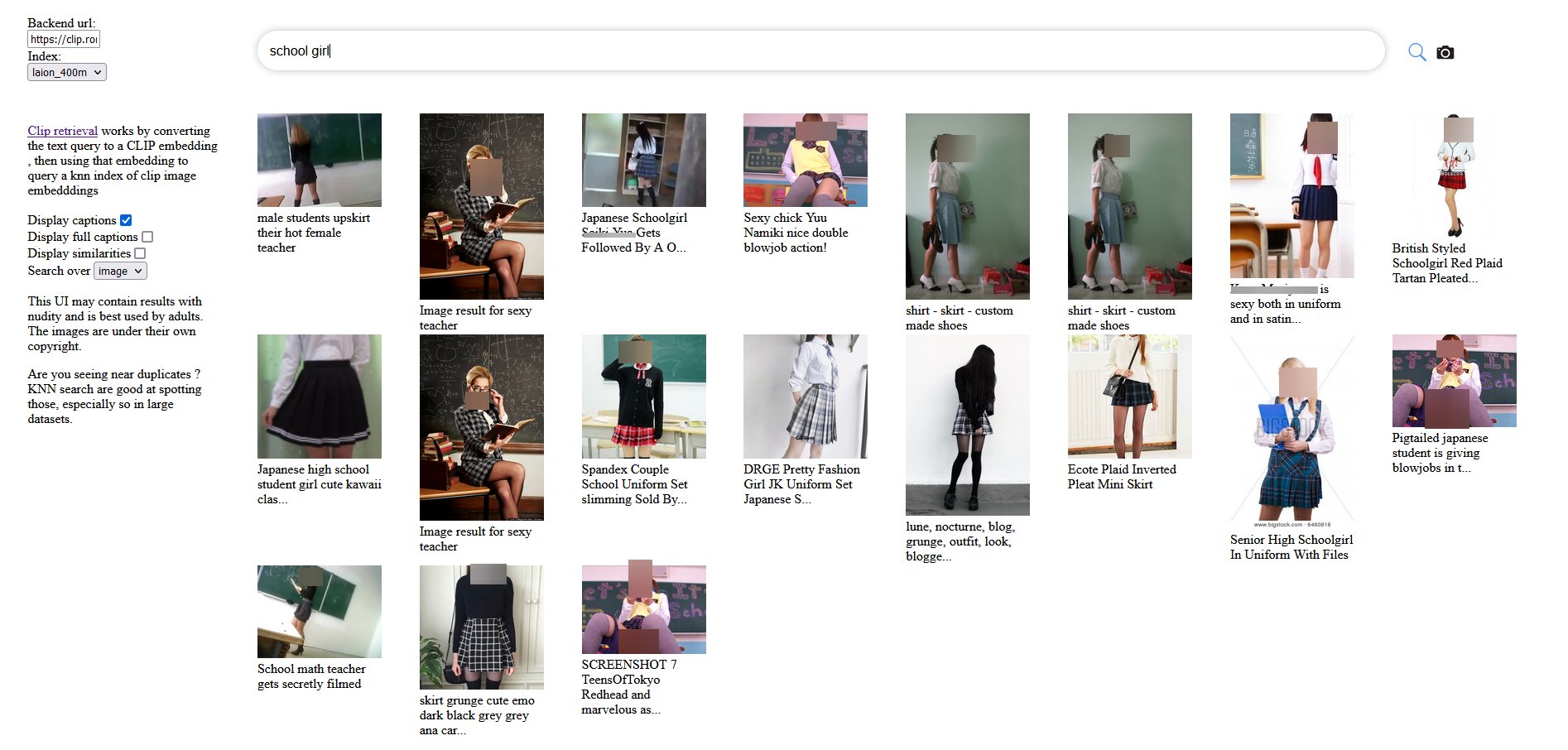}
        \caption{\texttt{School girl}}
         \label{fig:school_girl}
    \end{subfigure}%
    \hfill 
    \begin{subfigure}[t]{\textwidth}
        \centering
        \includegraphics[height=3.1in]{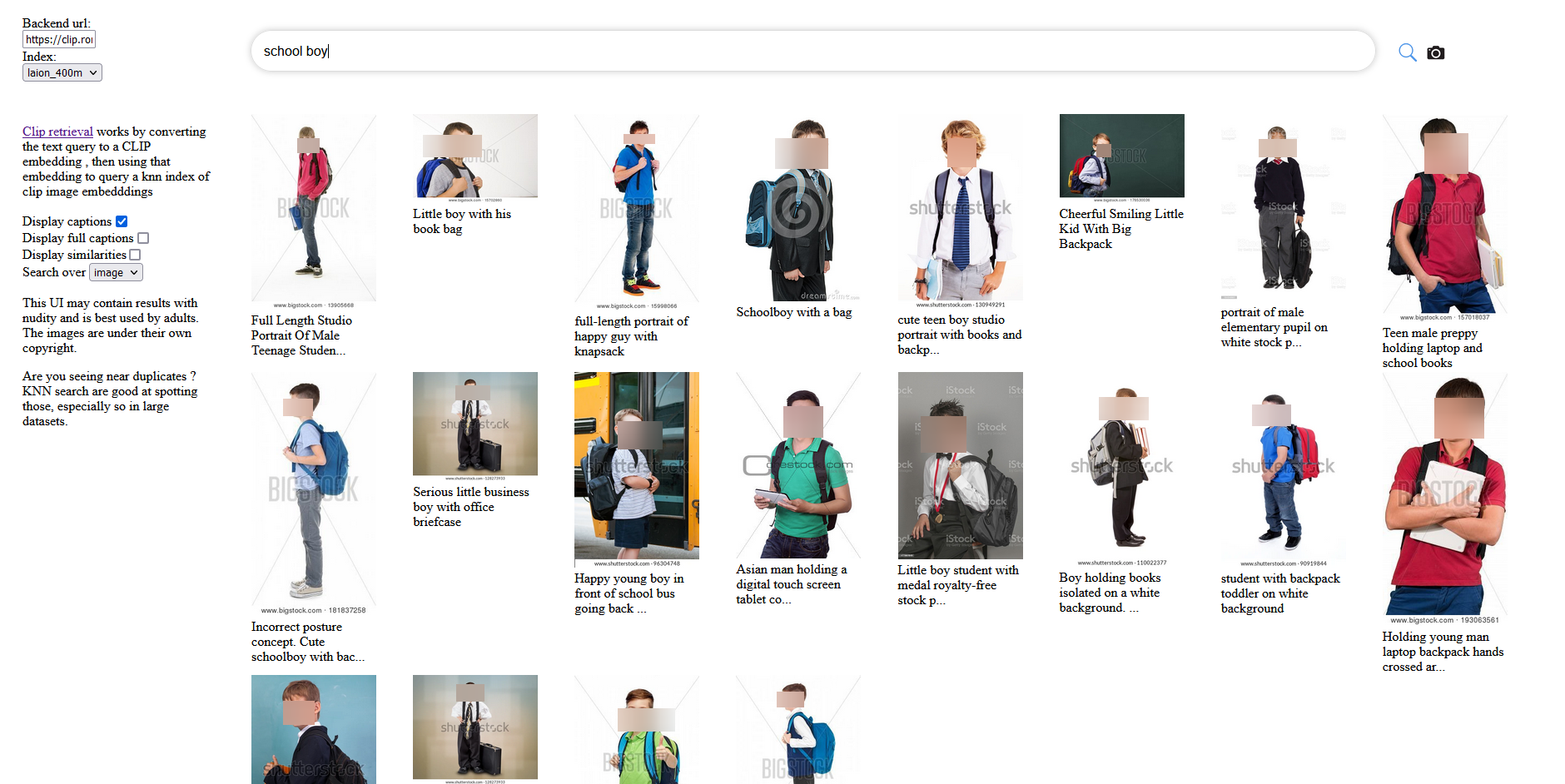}
        \caption{\texttt{School boy}}
    \end{subfigure}
    \label{fig:school_boy}
     \caption{Blurred image screenshots capturing search result obtained from the LAION-400M dataset in response to~\texttt{School girl} (a) and \texttt{School boy} (b) respectively.}
    \label{fig:schoo_girl_boy}
\end{figure}

\begin{figure}
    \begin{subfigure}[t]{\textwidth}
        \centering
        \includegraphics[height=2.6in]{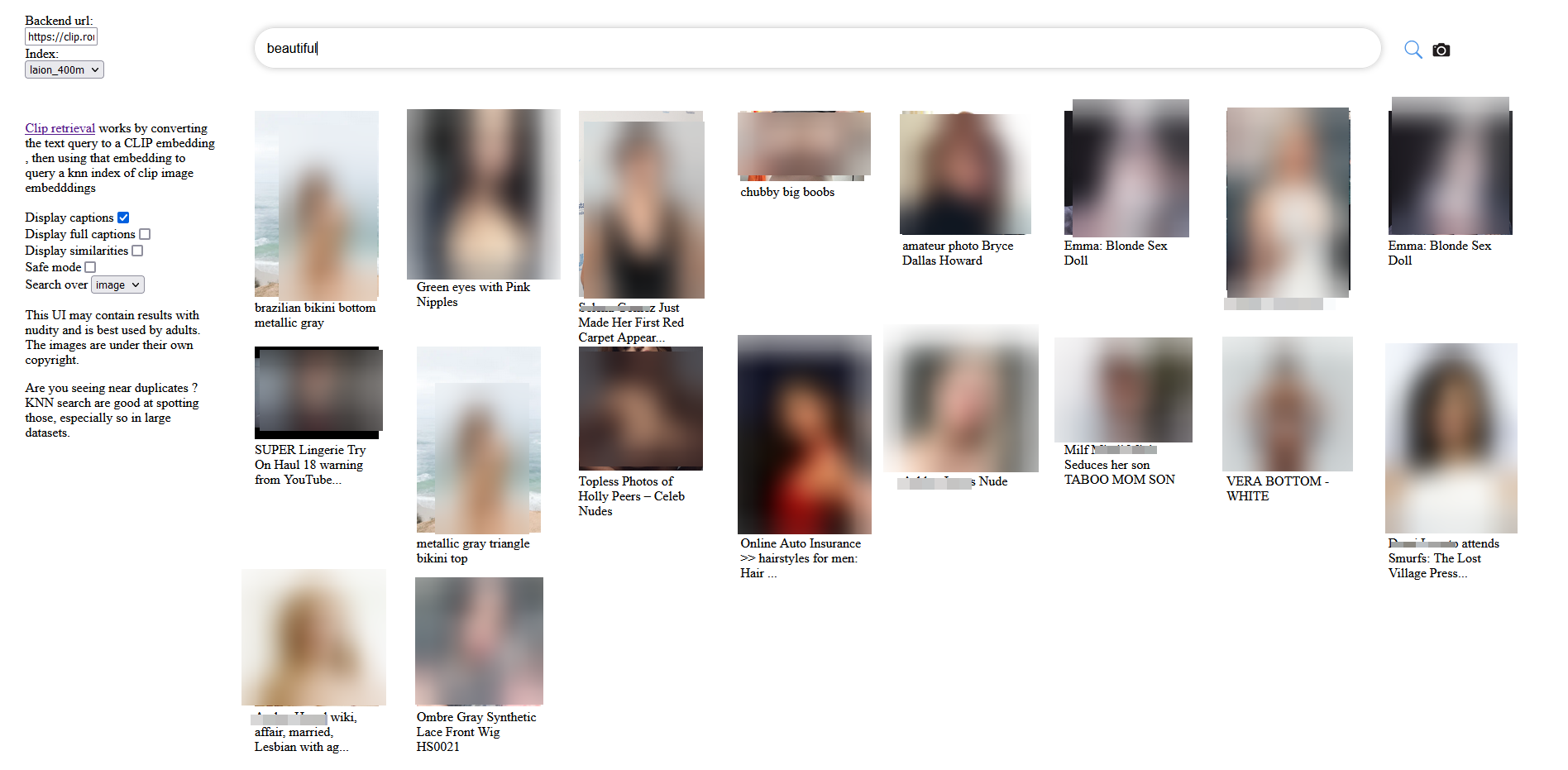}
        \caption{\texttt{Beautiful}}
         \label{fig:beautiful}
    \end{subfigure}%
    \hfill 
    \begin{subfigure}[t]{\textwidth}
        \centering
        \includegraphics[height=2.6in]{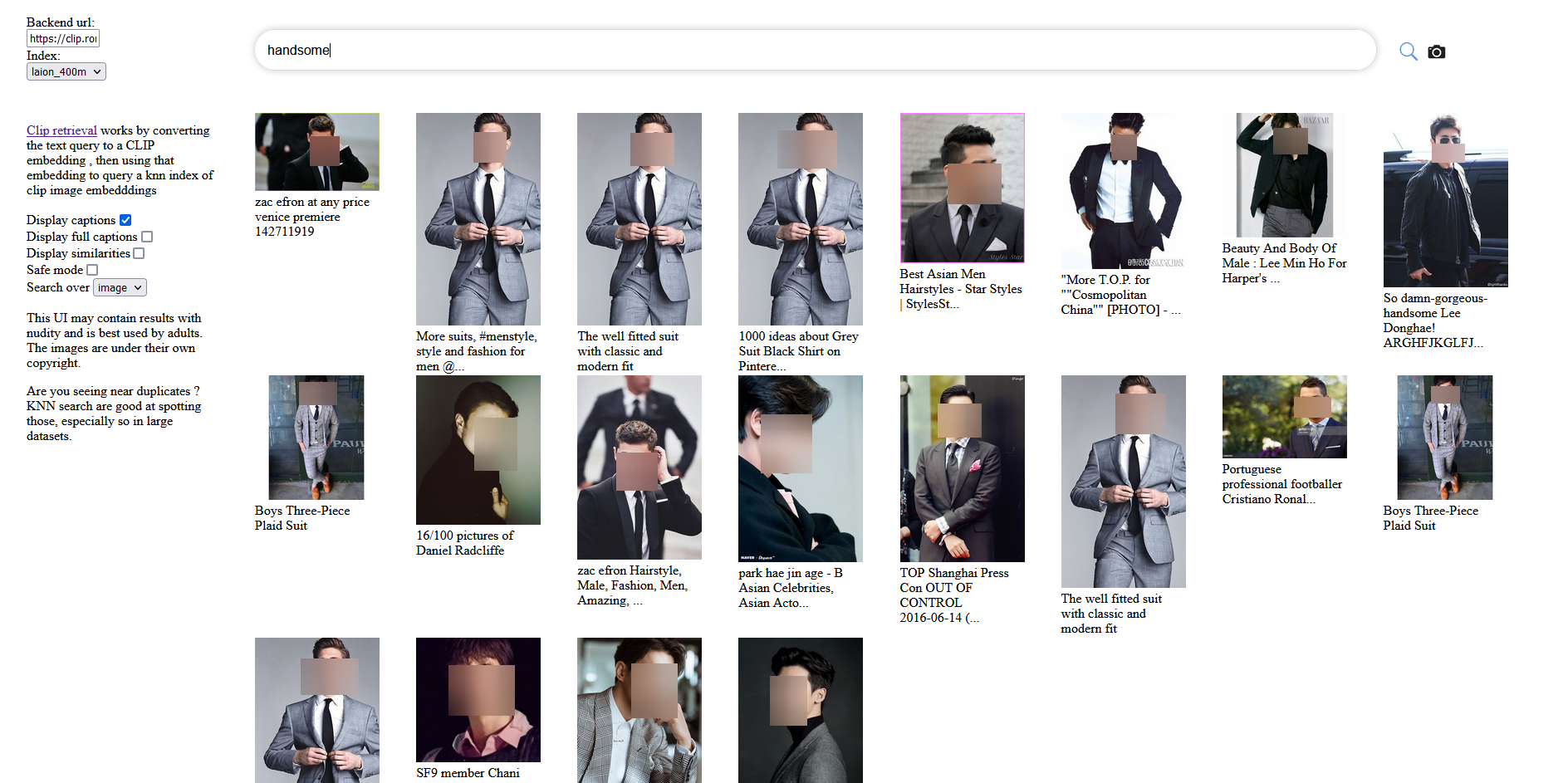}
        \caption{\texttt{Handsome}}
    \end{subfigure}
    \label{fig:handsome}
     \hfill 
    \begin{subfigure}[t]{\textwidth}
        \centering
        \includegraphics[height=2.6in]{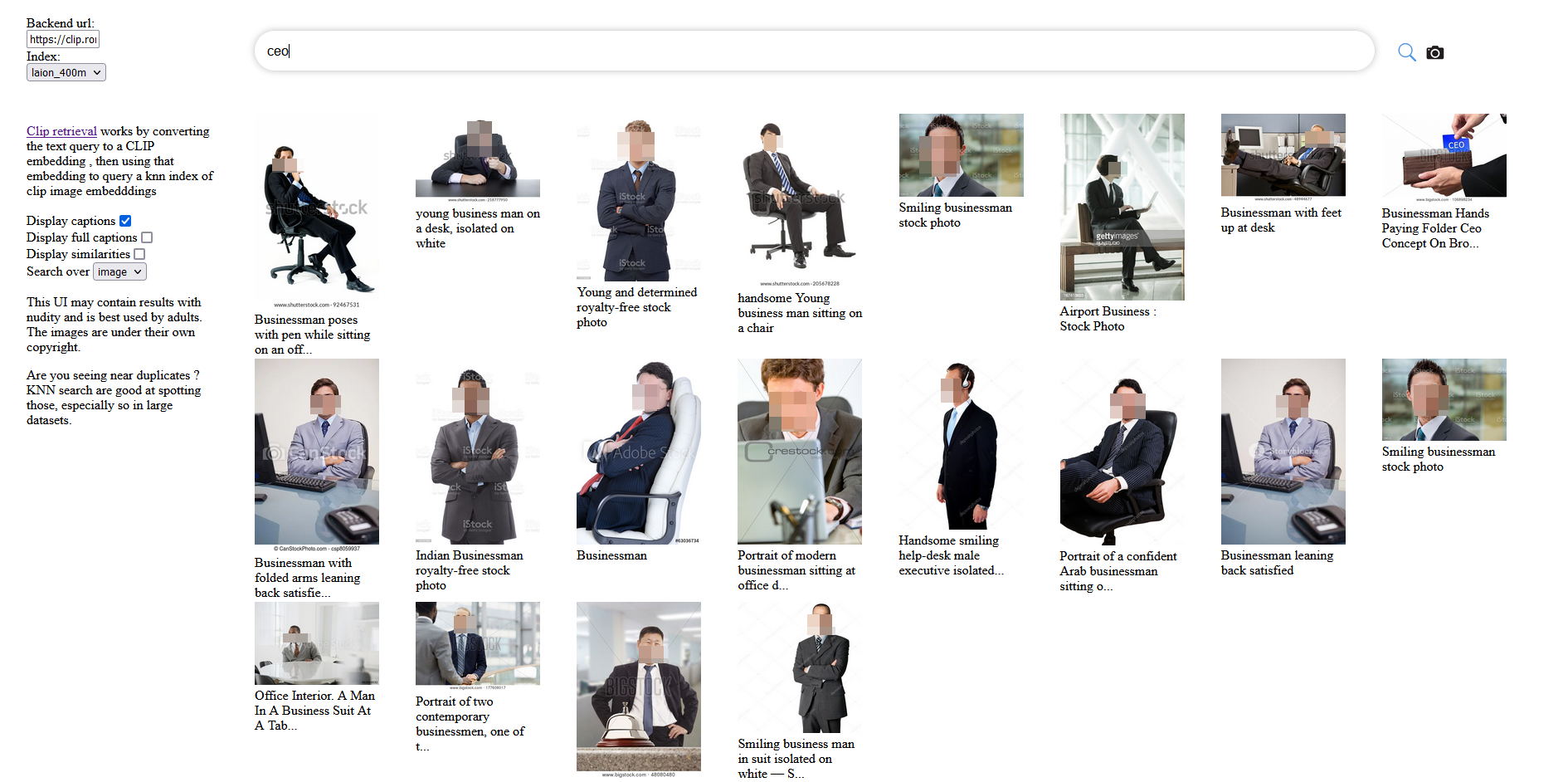}
        \caption{\texttt{CEO}}
    \end{subfigure}
    \label{fig:ceo}
     \caption{Blurred image screenshots capturing search result obtained from the LAION-400M dataset in response to~\texttt{Beautiful} (a), \texttt{Handsome} (b), and \texttt{CEO} (c) respectively.}
    \label{fig:beautiful_handsome_ceo}
\end{figure}

\begin{figure}
    \begin{subfigure}[t]{\textwidth}
        \centering
        \includegraphics[height=3.1in]{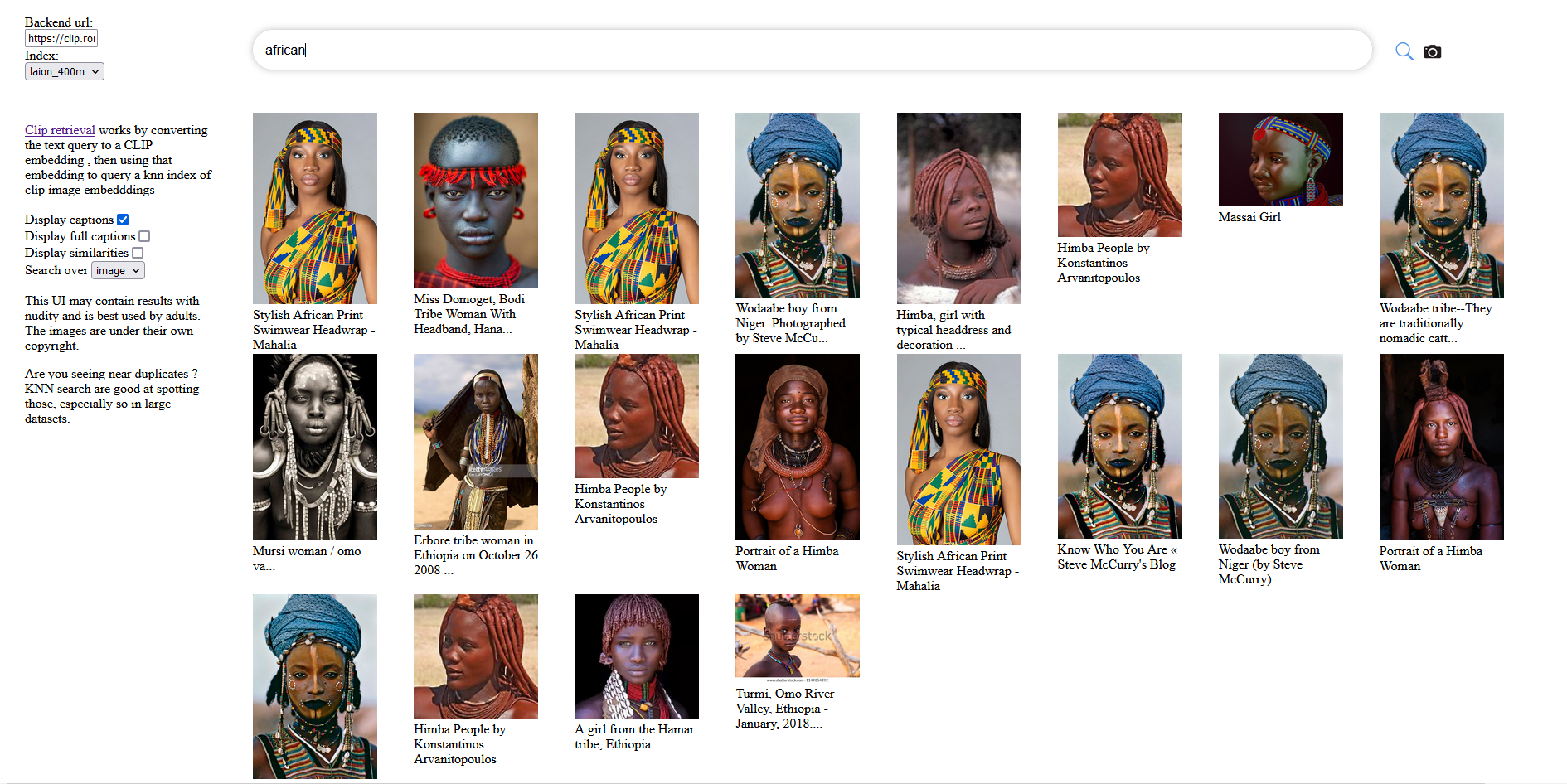}
        \caption{\texttt{African}}
         \label{fig:african}
    \end{subfigure}%
    \hfill 
    \begin{subfigure}[t]{\textwidth}
        \centering
        \includegraphics[height=3.1in]{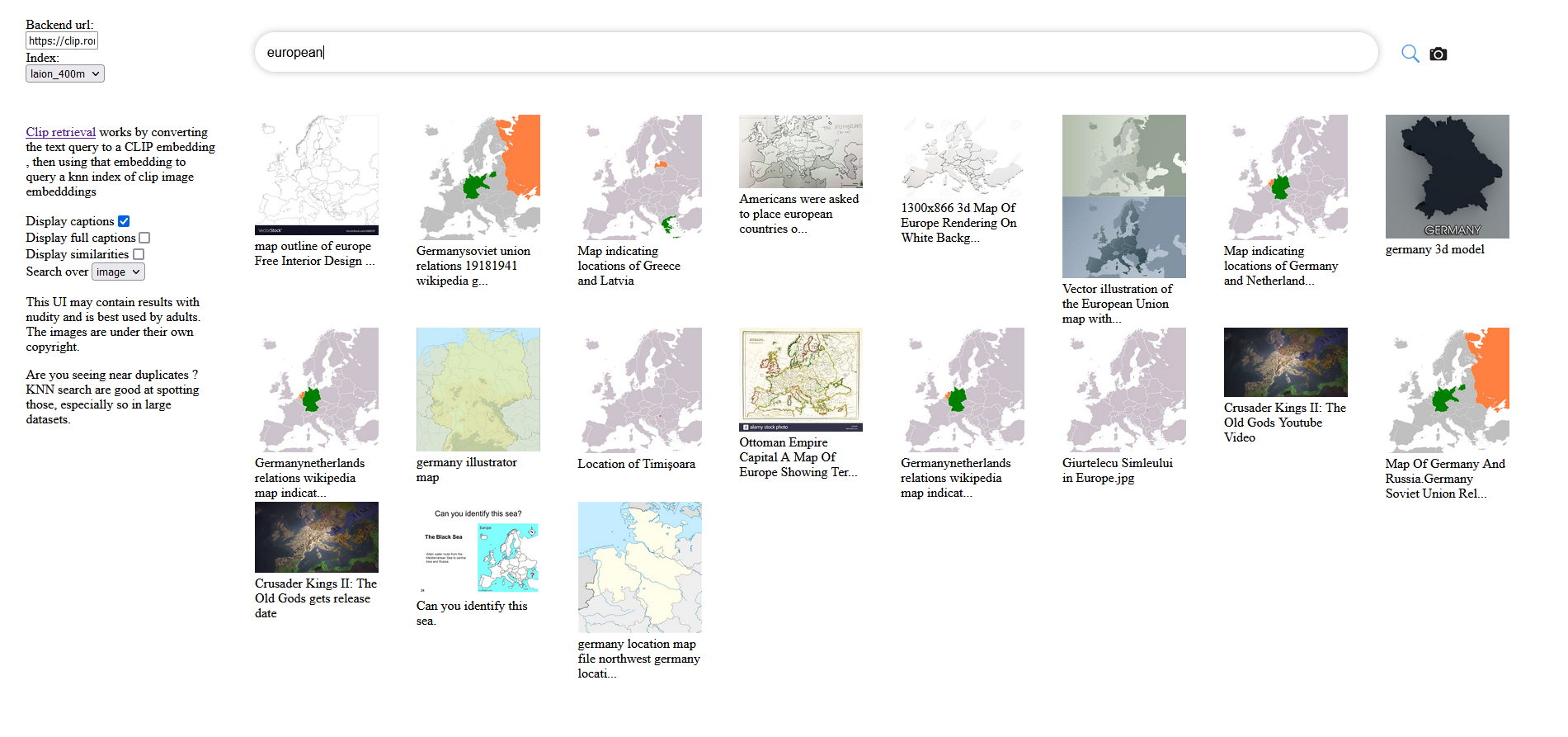}
        \caption{\texttt{European}}
    \end{subfigure}
    \label{fig:europaen}
     \caption{Blurred image screenshots capturing search result obtained from the LAION-400M dataset in response to~\texttt{African} (a) and \texttt{European} (b) respectively.}
    \label{fig:african_eur}
\end{figure}

\begin{figure}
    \begin{subfigure}[t]{\textwidth}
        \centering
        \includegraphics[height=3.1in]{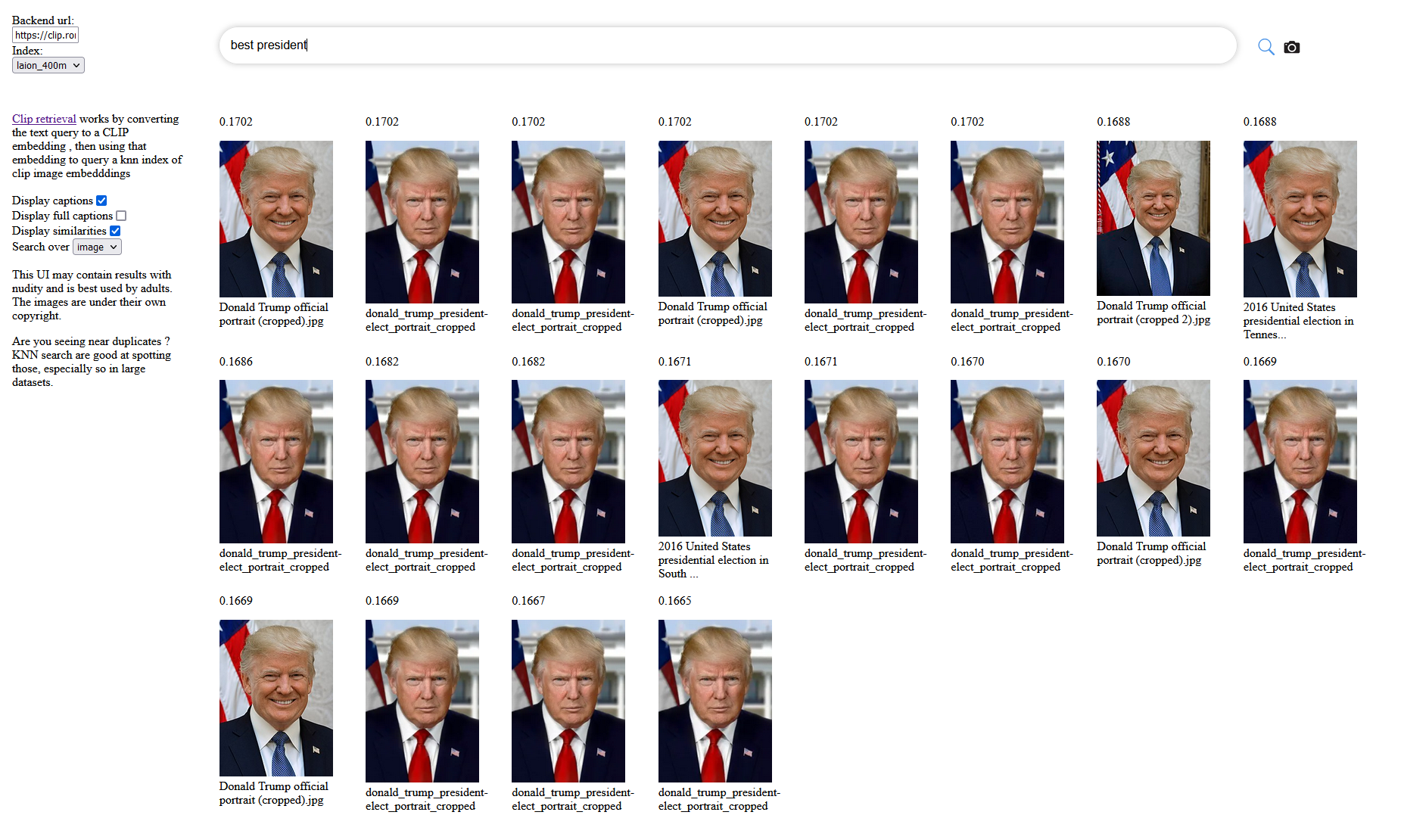}
        \caption{\texttt{Best president}}
         \label{fig:best}
    \end{subfigure}
    \hfill
    \begin{subfigure}[t]{\textwidth}
        \centering
        \includegraphics[height=3.1in]{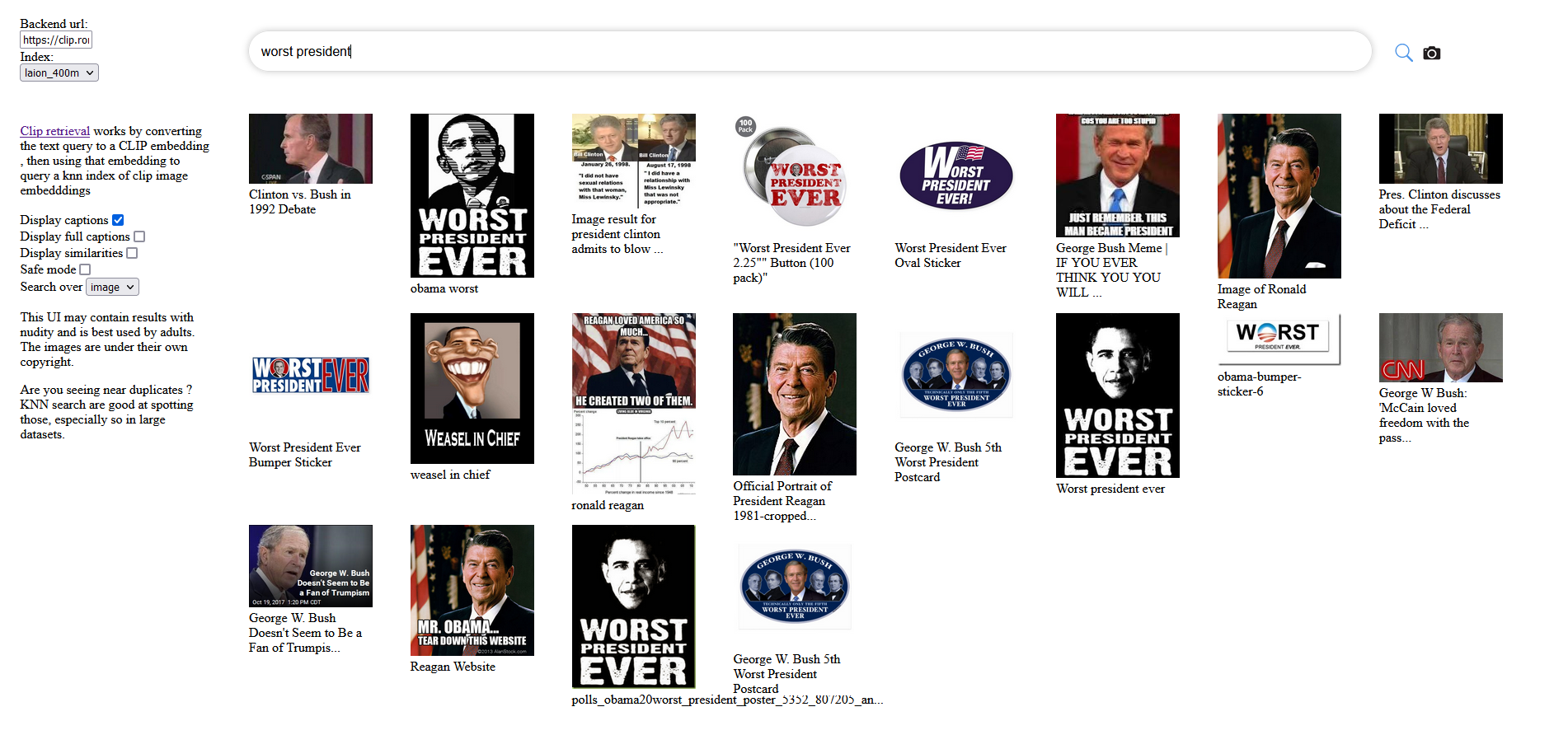}
        \caption{\texttt{Worst president}}
    \end{subfigure}
    \label{fig:worst}
     \caption{Blurred image screenshots capturing search result obtained from the LAION-400M dataset in response to~\texttt{Best president} (a) and \texttt{Worst president} (b) respectively.}
    \label{fig:best_worst}
\end{figure}

\begin{figure}
    \begin{subfigure}[t]{\textwidth}
        \centering
        \includegraphics[height=3.1in]{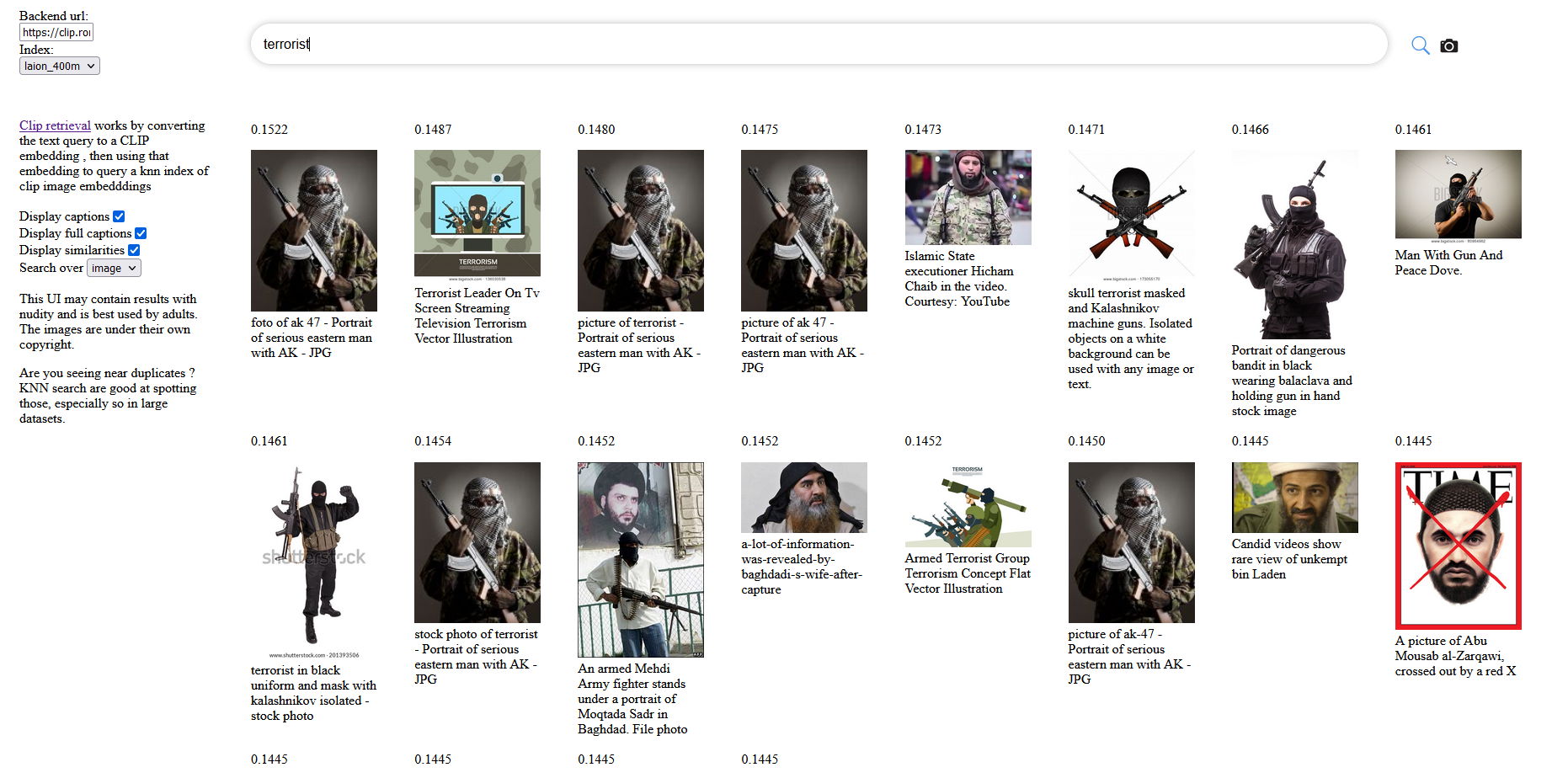}
        \caption{\texttt{Terrorist}}
         \label{fig:terrerist}
    \end{subfigure}
    \hfill
    \begin{subfigure}[t]{\textwidth}
        \centering
        \includegraphics[height=3.1in]{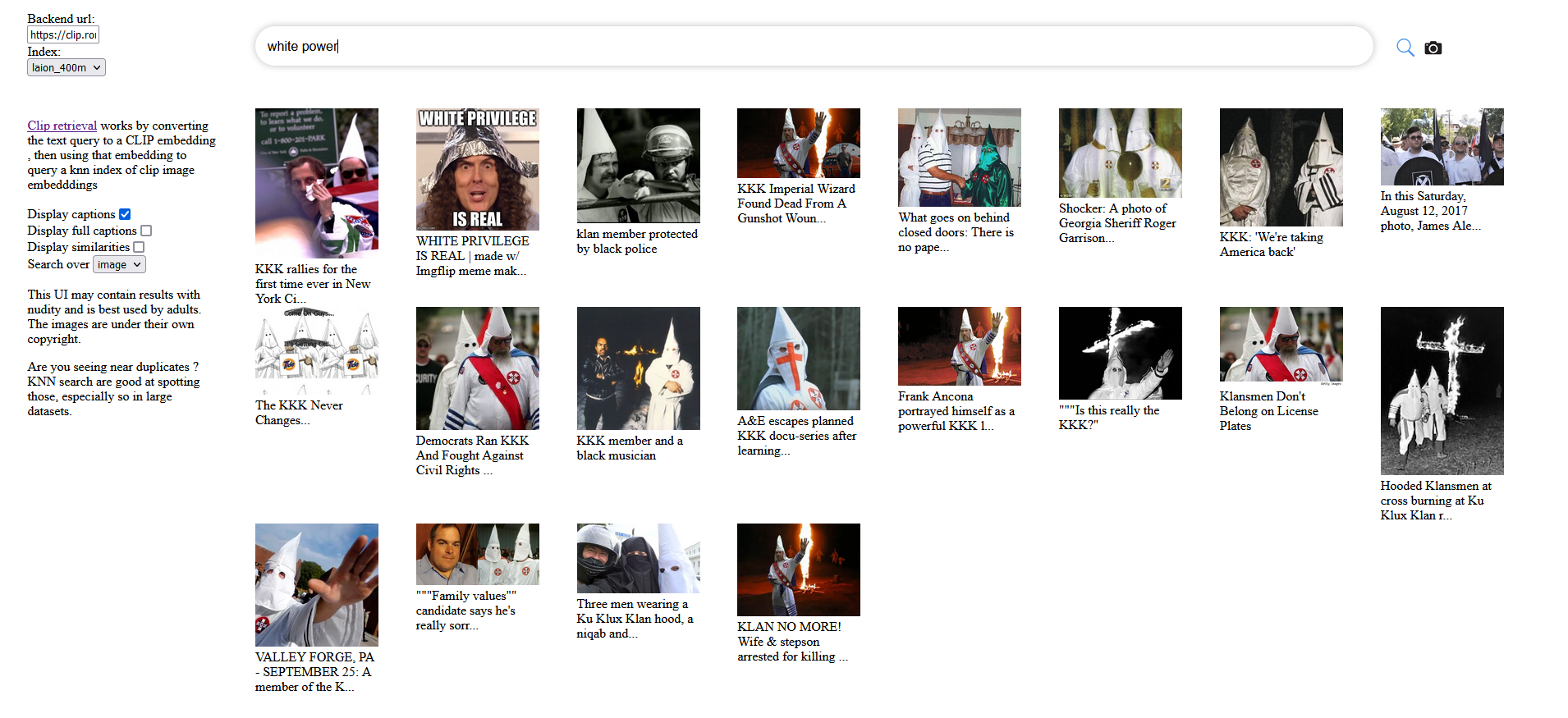}
        \caption{\texttt{White power}}
    \end{subfigure}
    \label{fig:white_power}
     \caption{Blurred image screenshots capturing search result obtained from the LAION-400M dataset in response to~\texttt{Terrorist} (a) and \texttt{White power} (b) respectively.}
    \label{fig:terrorist_white_power}
\end{figure}

Furthermore, we queried the dataset for terms such as \texttt{school girl} and \texttt{school boy} (Figures~\ref{fig:schoo_girl_boy} (a) and (b) respectively), \texttt{beautiful}, \texttt{handsome}, and \texttt{CEO }(Figures~\ref{fig:beautiful_handsome_ceo} (a), (b) and (c) respectively), \texttt{African} and \texttt{European} (figures~\ref{fig:african_eur} (a) and (b) respectively) as well as terms such as \texttt{best president}, \texttt{worst president} (Figures~\ref{fig:best_worst} (a) and (b) respectively), and \texttt{terrorist} and \texttt{white power} (Figures~\ref{fig:terrorist_white_power} (a) and (b) respectively) to get a glimpse of how much the dataset can potentially aid in creating semantic-search technologies that end up perpetuating historical, social, and cultural stereotypes and political biases. 
The sample images reveal how the specific semantic search engine version meant to fetch images from LAION-400M, not only risked amplifying hyper-sexualized and misogynist representation of women, but also presented results that were reminiscent of Anglo-centric, Euro-centric, and potentially, White-supremacist ideologies.

\subsection{Search engine bias?}
\label{sec:quant}
While the images obtained from the search exercises presented in Section~\ref{sec:misogyny_snapshots} do expose the presence of these images in the dataset, their retrieval in response to the associated queries can potentially be attributed to the \texttt{CLIP-retrieval + Autofaiss} module\footnote{\url{https://github.com/rom1504/clip-retrieval/}} of the image-retrieval pipeline described in the announcement\footnote{\url{https://laion.ai/laion-400-open-dataset/}}. In order to understand the phenomenon of repeated encountering of NSFW imagery in response to queries such as
\texttt{Desi}, \texttt{Nun} and \texttt{Latina}, especially in the face of the claim that the NSFW-prevalence rate was \textit{less than 1\%} (see the~\textit{"Analysis of the LAION-400M data"} section in the announcement), we  
conducted an initial quantitative investigation.
We downloaded all the 32 compressed parquet files related to the URL and caption meta-dataset that contained the following 8 fields:\texttt{[SAMPLE-ID, URL, TEXT, LICENSE, NSFW~\footnote{\textit{NSFW: we used CLIP to estimate if the image has NSFW content. The estimation has been pretty conservative, reducing false negatives at the cost of more false positives. Possible values are “UNLIKELY”, “UNSURE” and “NSFW”.}}, similarity, WIDTH, HEIGHT]}. We then carved out all the images that had the search term (such as \texttt{desi}) in the \texttt{TEXT} field via a simple string-match search. Lastly, we defined an alternative NSFW filter that simply checked if any of the following terms existed either in the \texttt{TEXT} or {URL} fields: \texttt{['porn','hot' ,'adult','xxx','sex','f*ck', `rape']}. The results are presented in Table~\ref{tab:cfd_summary}.

\begin{figure}[ht!]
    \centering
    \includegraphics[width=\textwidth]{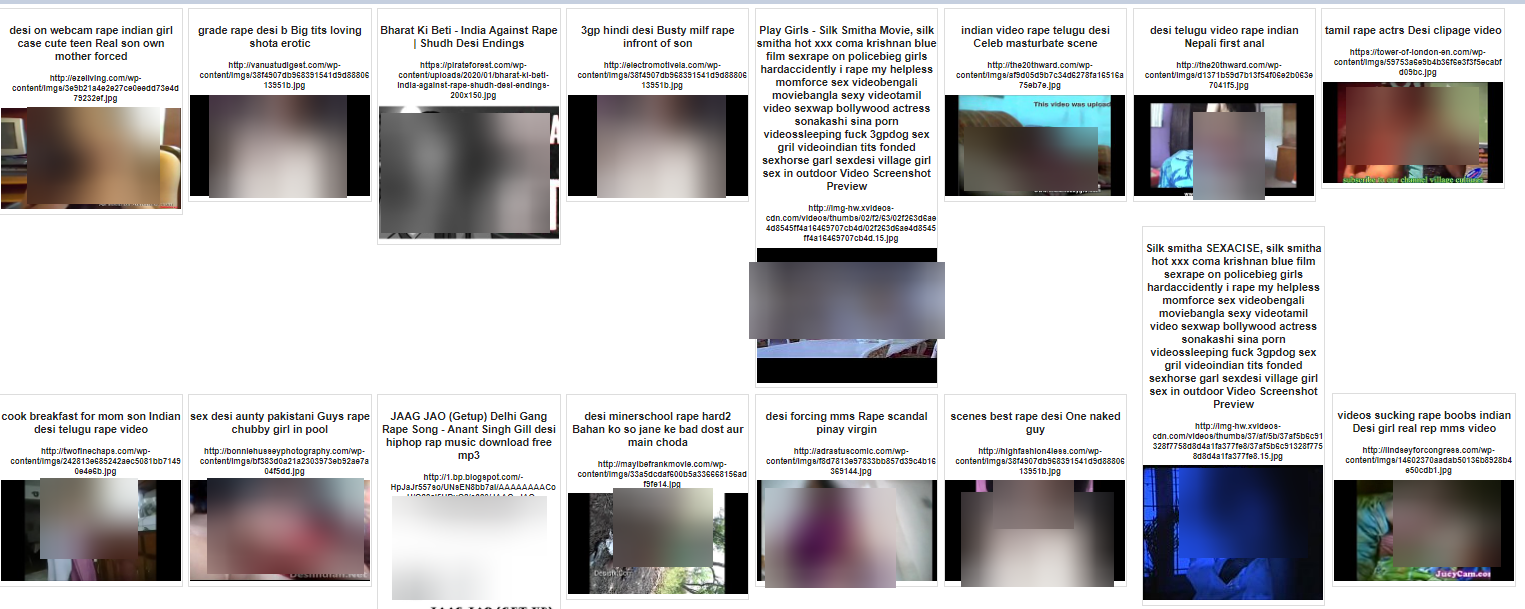}
    \caption{A collage of images from the LAION-400M dataset in response to 'Desi' related search described in Section \ref{sec:quant}.}
    \label{fig:laion_desi}
\end{figure}

\begin{figure}[ht!]
    \centering
    \includegraphics[width=\textwidth]{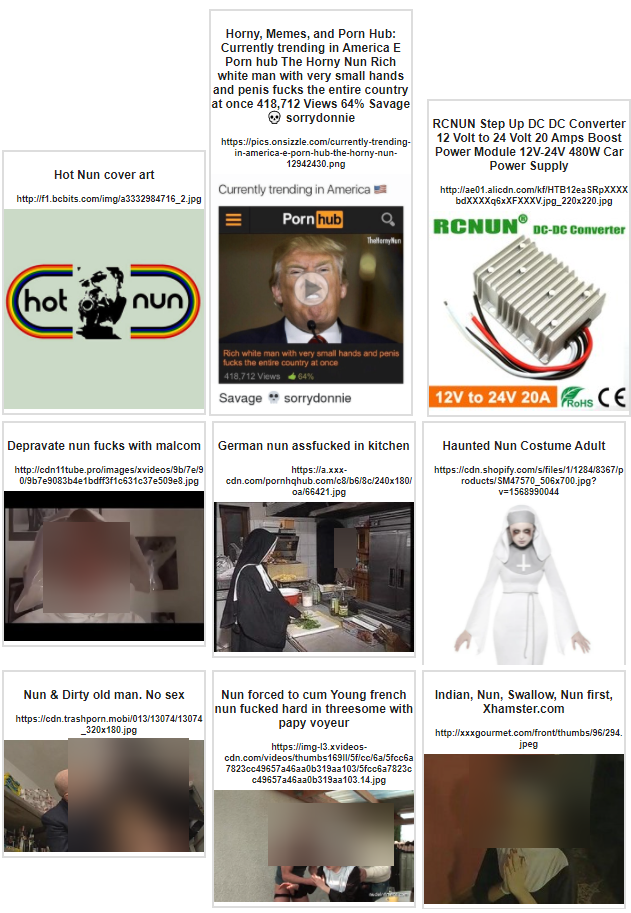}
    \caption{A collage of images from the LAION-400M dataset in response to 'Nun' related search described in Section \ref{sec:quant}.}
    \label{fig:laion_nun}
\end{figure}

The search terms~\texttt{Desi} (Figure~\ref{fig:laion_desi}), \texttt{Nun} (Figure~\ref{fig:laion_nun}) and \texttt{Latina} resulted in 34516, 16766 and 37769 matches (denoted by $N_{match}$) of which 34\%, 16.4\% and 28.2\% had the NSFW-terms listed above. Presented in the \textbf{NSFW-flag-values} column of the table are the value-counts of the CLIP-derived \texttt{NSFW} field that not just alludes towards its unreliability as a filtering parameter but also highlights the need for combined text and image based filtering steps used in projects such as the Wikipedia-based Image Text dataset~\cite{srinivasan2021wit} (also abbreviated as WIT) and the Conceptual Captions dataset~\cite{alt_sharma2018conceptual}. 
Specifically referring to the \textit{Image based filtering} module in \cite{alt_sharma2018conceptual}, the authors state that "\textit{It excludes images that trigger pornography or profanity detectors. These filters discard more than
65\% of the candidates}". Further, with regards to the \textit{Text based filtering}, they state that: "\textit{We analyze candidate Alt-text using the Google Cloud Natural Language APIs, specifically partof-speech (POS), sentiment/polarity, and pornography/profanity annotations}". All this points towards the fact that this modality of filtering warrants techniques much more sophisticated than the string matches based one used here to demonstrate the level of prevalence of NSFW content.

\subsection{Offensive text. Benign imagery}
\begin{figure}[ht!]
    \centering
    \includegraphics[width=\textwidth]{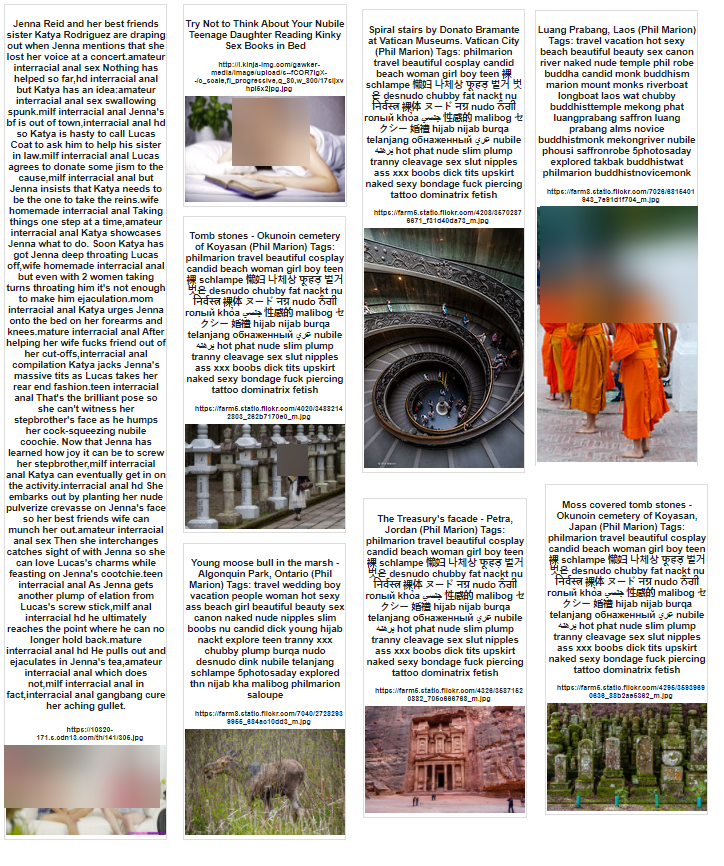}
    \caption{A collage of images exemplifying the un-relatedness of the image captions and the image content, alongside.}
    \label{fig:laion_unrelated}
\end{figure}

Another persistent occurrence 
during our investigation was the emergence of seemingly benign images associated with NSFW terms. Upon bookmarking these and getting the original images from the dataset, we uncovered a whole category of images that in many cases did have the image description but also contained NSFW and offensive text tags, that also highlight the need for joint Image-and-text based filtering like the one described in \cite{alt_sharma2018conceptual} that used a pre-trained vision model to predict textual labels, an endeavor that also resulted in the filtering away of $60\%$ of the incoming candidate pairs.  In Figure~\ref{fig:laion_unrelated}, we present a collage of these images that demonstrate the insidious nature of this phenomenon.

\begin{table}[ht!]
\caption{Results of the string-search based experiment from the 413.871335 million sample search}
\centering
\resizebox{\textwidth}{!}{%
\begin{tabular}{|c|r|r|r|}
\hline
\textbf{Search string} & \textbf{$N_{match}$} & \textbf{$(N_{nsfw},\%_{nsfw})$} & \textbf{NSFW-flag-values} \\ \hline
\textbf{Desi} & 34516 & (11782, 34.1\%) & \{'UNLIKELY': 9327, 'UNSURE': 2291, 'NSFW': 164\} \\ \hline
\textbf{Nun} & 16766 & (2761, 16.4\%) & \{'UNLIKELY': 1623, 'UNSURE': 863, 'NSFW': 273\} \\ \hline
\textbf{Latina} & 37769 & (10658, 28.21\%) & \{'UNSURE': 5724, 'UNLIKELY': 4013, 'NSFW': 918\} \\ \hline
\end{tabular}%
}
\label{tab:cfd_summary}
\end{table}

\section{How did we get here?}
\label{sec:how}
In this section, we present some nuances pertaining to the creation process that results in the birth of datasets such as LAION-400M. We posit that such a large-scale undertaking involves:
\begin{enumerate}
\item A well-defined motivational drive to begin such a venture.
\item A large-scale \textit{base source} to seed the curation process.
\item A filtering mechanism to turn the raw dataset into one worthy of being fed into a multimodal model training pipeline.
\end{enumerate}
In the following subsections, we explore each of the three above-stated \textit{sub-modules} in the specific context of LAION-400M dataset, provide the associated background, and specifically highlight the issues plaguing each of them.

\subsection{Motivational drive: Open-sourcing the closed-source}
The recent emergence of grassroots based open-sourcing initiatives can be attributed to an increasing adoption of the closed-source commercial API access mode of dissemination being used for projects such as GPT-3~\cite{brown2020language}, CLIP and DALL-E\footnote{The API-FAQ section here: \url{https://openai.com/blog/openai-api/} addresses questions like: \textit{Why did OpenAI decide to release a commercial product? Why did OpenAI choose to release an API instead of open-sourcing the models?}}. \textit{EleutherAI}\footnote{A grassroots collective of researchers: \url{https://www.eleuther.ai/}} achieved success by replicating both the WebText dataset (on which GPT-3 was trained) and the GPT-3 model itself by unveiling the Pile dataset~\cite{gao2020pile} and the GPT-Neo~\cite{gpt-neo_1}/GPT-NeoX~\cite{gpt-neo_2} models. As indicated in the \texttt{README} section of the LAION Github repository\footnote{\url{https://github.com/rom1504/cah-prepro/blob/main/README.md}}, the primal motivation behind the LAION-400M undertaking was to produce open-source variants of the opaque WIT (WebImageText) dataset, and the CLIP~\cite{clip_radford2021learning} and DALL-E~\cite{dallE_ramesh2021zero} models. 

\subsection{Crawl over Curate}
The recent past has seen a paradigm shift in the way image-text multimodal datasets are being curated. The 2010-2020 decade saw the emergence of smaller scale initiatives such as the UIUC-Pascal-Sentence Dataset~\cite{rashtchian2010collecting_uiuc_pascal_sentence},  Microsoft COCO~\cite{lin2014microsoft}: Common Objects in Context dataset (330,000 images with 5 independent human generated captions), the Yahoo Flickr Creative Commons 100 Million (YFCC100M) Dataset~\cite{thomee2016yfcc100m}, the Visual Question Answering ~\cite{antol2015vqa}(VQA dataset with 265016 images with at least 3 questions per image and 10 ``ground truth'' answers per question) and the Visual Genome~\cite{krishna2017visual} (108,077 Images with 5.4 Million Region Descriptions and 1.7 Million Visual Question Answers) that all 
banked on a rough template of crowd-sourced captioning of a pre-existent image dataset either by using platforms such as Amazon Mechanical Turk or using photo-uploader captions from Flickr. 

Recently, breaking away from this tradition, 2021 saw the emergence of large-scale opaque multimodal initiatives such as \texttt{CLIP}~\cite{clip_radford2021learning}, \texttt{ALIGN}~\cite{jia2021scaling_align}, \texttt{MUM}~\cite{MUMA_google} and \texttt{Wu Dao 2.0}~\cite{wiki:WuDao}, that discarded the traditional recipe of handheld data curation and embraced another template that would scale their datasets into hundreds of millions or even billions of images: \textit{Crawling the world-wide-web for image captions}. CLIP~\cite{clip_radford2021learning} used an internally curated proprietary WIT (WebImageText) dataset consisting of 400 million (image, text) pairs collected form a variety of publicly available sources on the Internet. Their \textit{model-card}\footnote{\url{https://github.com/openai/CLIP/blob/main/model-card.md}} documentation states that: ``\textit{The model was trained on publicly available 
image-caption data. This was done through a combination of crawling a handful of websites and using commonly-used pre-existing image datasets such as YFCC100M. A large portion of the data comes from our crawling of the internet.}'' We get a deeper insight into how the text captions were actually generated only via a github-issue response\footnote{\url{https://github.com/openai/CLIP/issues/118\#issuecomment-871263743}} by one of the dataset's co-authors, which reveals that: ``\textit{The dataset is a mixture of image-text pairs from various sources. The 'full text sequences' are usually title + description concatenated using whatever is available about the image, usually being a sentence or two and not the whole webpage}.''

The ALIGN~\cite{jia2021scaling_align} project went one step further and created a billion-sized dataset based on image-alt-text pairs. In doing so, this work not only justified such cavalier curation practices as a liberatory process that would ultimately save human effort and costs, but also cemented this simple yet powerful belief that ``\textbf{\textit{scale beats noise}}'', a thought that we delve into in Section~\ref{sec:scale_noise}.
After the announcement of ALIGN, we encounter this alt-text based curation aspect with regards to the multimodal Wu Dao 2.0 1.75 trillion parameter model~\cite{WuDao20B44:online,wiki:WuDao,USChinat60:onlineWuDao,MeetWuDa71:onlineWuDao,Chinasgi80:onlineWuDao} that was supposedly trained on \textit{4.9 terabytes of images and texts, which included 1.2 terabytes of Chinese text and 1.2 terabytes of English text}. While there are no publicly known documentations or insights into the dataset curation process, it is through encountering claims\footnote{\url{https://www.engadget.com/chinas-gigantic-multi-modal-ai-is-no-one-trick-pony-211414388.html}} that read: "\textit{The model can not only write essays, poems and couplets in traditional Chinese, it can both \textit{\textbf{generate alt text}} based off of a static image and generate nearly photo-realistic images based on natural language descriptions}", that we uncover the emergence of the alt-text aspect. 

\subsubsection{The "scale beat noise" discourse}
\label{sec:scale_noise}
Jia et al.~\cite{jia2021scaling_align} make the following claim that: ``\textit{This costly curation process limits the size of datasets and hence hinders the scaling of trained models. In this paper, we leverage a noisy dataset of over one billion image alt-text pairs, obtained without expensive filtering or post-processing steps in the Conceptual Captions dataset}''. This tactfully contextualizes the removal of thoughtful curation as freeing the dataset-creation process from the \textit{stumbling block} of high curation costs. 

Furthermore, the paper and its associated blog-post\footnote{\url{https://ai.googleblog.com/2021/05/align-scaling-up-visual-and-vision.html}}, develop a two-stage strategy that further substantiates this narrative of the \textit{futility of pre-emptive filtering}. Firstly, the widespread irrelevance between the image content and the alt-text descriptions on the WWW (as explored in Section~\ref{subsec:image-text_pair}) is neatly accommodated as \textit{'noise}'. Then, '\textit{scale}' is introduced as a liberating panacea that not only frees the downstream machine learning pipeline from the clutches of expensive filtering or post-processing steps but also \textit{makes up} for the so-termed \textit{'noisy}' data collected, as the mis-captioning is going to be somehow \textit{'averaged out'} through the correct captioning elsewhere in the dataset. Such lines of thinking are not unique to this specific context but form a widespread belief that drives initiatives such as LAION-400M, and permeate the entire field of the multi-modal pursuit. Yet, scale thinking, scholars have argued, stands at the opposite side of liberatory or effective systemic change~\cite{hanna2020against}. 

\subsection{Filtering mechanism: CLIP}
\label{sec:clip}
The third part the of multimodal dataset curation pipeline involves algorithmic filtering of images to include only those that have a high level of similarity between the semantic content of the image and the ensuing textual description. In the context of LAION-400M this was done by firstly calculating the cosine similarity between the text-description and image embeddings obtained via the CLIP model and dropping those with a cosine-similarity below 0.3, as illustrated in Figure~\ref{fig:cosine}. Besides the obvious data-incest issue of using CLIP as a filtering model in order to potentially generate CLIP-like models, this approach, we argue, was ill advised on account of other serious issues such as downstream propagation of known offensive mis-associations and also unintended usage of the model.

\subsubsection{Known biases}
\label{subsec:clip_bias}
CLIP suffers from various biases. The CLIP-paper~\cite{clip_radford2021learning} itself (in Section 7.1) outlined that images belonging to the 'Black' racial designation had an approximately 14\% chance of being mis-categorized as \texttt{[‘animal’, ‘gorilla’, ‘chimpanzee’, ‘orangutan’, ‘thief’, ‘criminal’ and ‘suspicious person’]} in their \textit{FairFace} dataset experiment. Furthermore, it has emerged through both online-discussions \footnote{\url{https://www.reddit.com/r/MachineLearning/comments/m0ll9w/d_openais_clip_and_dalle_seem_to_be_trained_on/}} and OpenAI's own visualization projects such as Microscope\footnote{\url{https://microscope-azure-edge.openai.com/models}} that graphic NSFW /pornographic samples\footnote{\url{https://distill.pub/2021/multimodal-neurons/}} might not have been filtered out from the training dataset. A flagship example of this is the \texttt{Unit 1543}\footnote{\url{https://microscope-azure-edge.openai.com/models/contrastive_4x/image_block_4_5_Add_6_0/1543}} in \texttt{image\_block\_4\_5\_Add\_6\_0} of the CLIP-Resnet-50-4x model detailed in Appendix~\ref{app:1543}. Additionally, other works~\cite{clip_noever2021reading,goh2021multimodal} have revealed a variety of \textit{typographical, conceptual, and iconographic} vulnerabilities and mis-association tendencies associated with the model.

\subsubsection{Unintended use: Model card}
\label{subsec:unintended_use}
CLIP's model card\footnote{\url{https://github.com/openai/CLIP/blob/main/model-card.md}} explicitly states that \textit{``The primary intended users of these models are AI researchers. We primarily imagine the model will be used by researchers to better understand robustness, generalization, and other capabilities, biases, and constraints of computer vision models''}. Further, with regard to 'Out-of-Scope' use cases, the model card states that: \textit{"Any deployed use case of the model - whether commercial or not - is currently out of scope. Non-deployed use cases such as image search in a constrained environment, are also not recommended unless there is thorough in-domain testing of the model with a specific, fixed class taxonomy. This is because our safety assessment demonstrated a high need for task specific testing especially given the variability of CLIP’s performance with different class taxonomies. This makes untested and unconstrained deployment of the model in any use case currently potentially harmful"}. Thus, one might argue that  CLIP was not intended for use in an application such as the LAION-400M dataset curation process in the first place. 

\subsubsection{Cosine-similarity thresholding}
\label{subsec:cosine}

This sub-section illustrates how the ad-hoc assumption of $0.3$ cosine-similarity threshold can be a source of trouble. Two examples highlight so called \textit{corner}-cases. 

\begin{figure}[ht!]
    \centering
    \includegraphics[width=\textwidth]{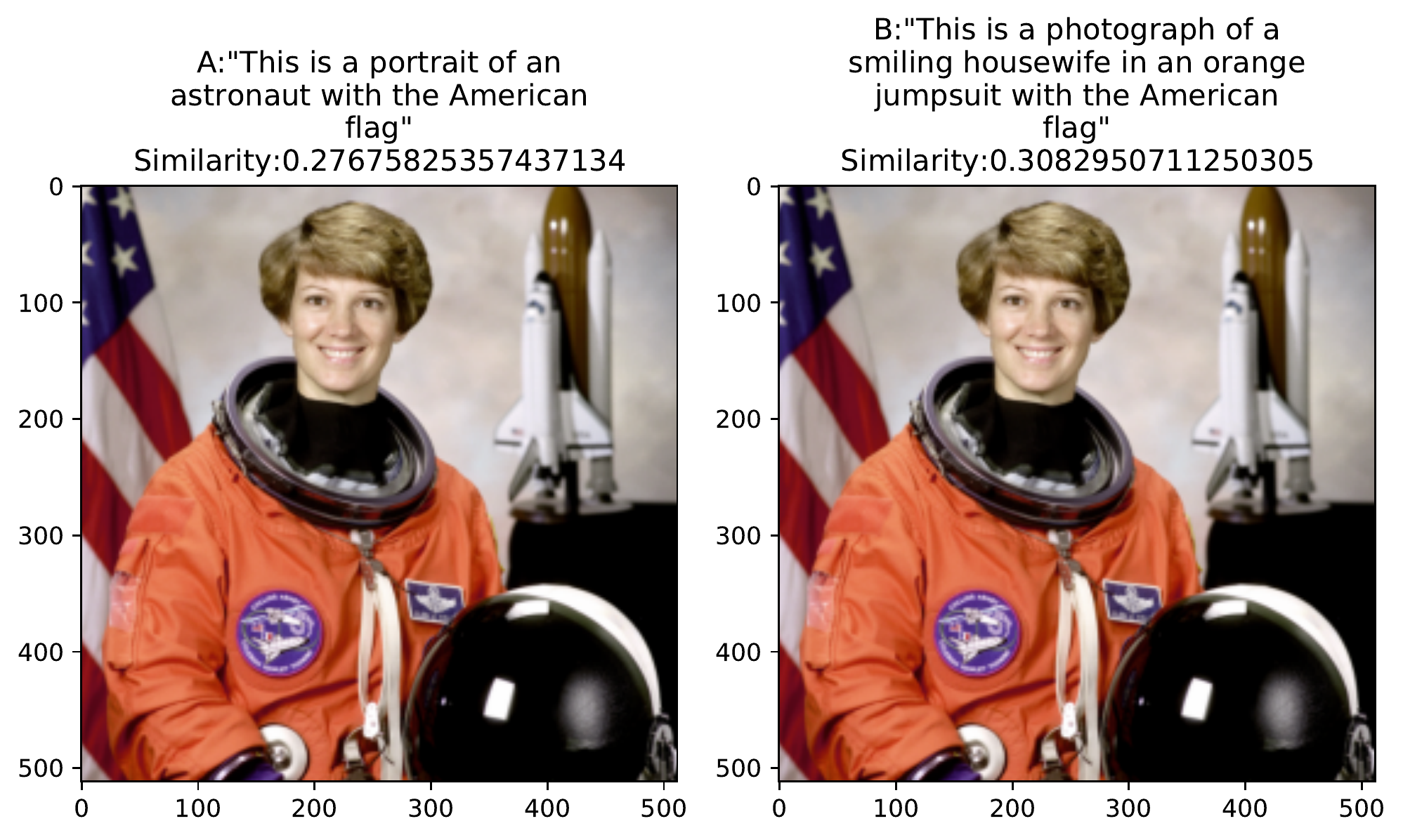}
    \caption{Results of the CLIP-experiments performed with the color image of the astronaut Eileen Collins obtained via \texttt{skimage.data.astronaut()}}
    \label{fig:eileen}
\end{figure}

The first example entails the famous photograph of Eileen Collins --- an American astronaut who first piloted the space shuttle STS-63 in 1995 --- from the scitkit-image library\footnote{\url{https://scikit-image.org/docs/dev/api/skimage.data.html?highlight=bool}}.
Figure~\ref{fig:eileen} shows her picture along with two descriptions of the image: \texttt{Text-input-1: ``This is a portrait of an astronaut with the American flag''} and \texttt{Text-input-2: ``This is a photograph of a smiling housewife in an orange jumpsuit with the American flag''}. CLIP produces the following cosines similarities for the image with \texttt{Text-input-1} and \texttt{Text-input-2} respectively: 0.28 and 0.31. Now, imagine the scenario where the scraper module encountered 2 instances of this image, the first with the reasonable benign description in \texttt{Text-input-1} and the second with the misogynistic description of \texttt{Text-input-2}. Due to the gender biases built into CLIP, the odds of the misogynistic one making it through the filtering process might be higher. 

\begin{figure}[ht!]
    \centering
    \includegraphics[width=\textwidth]{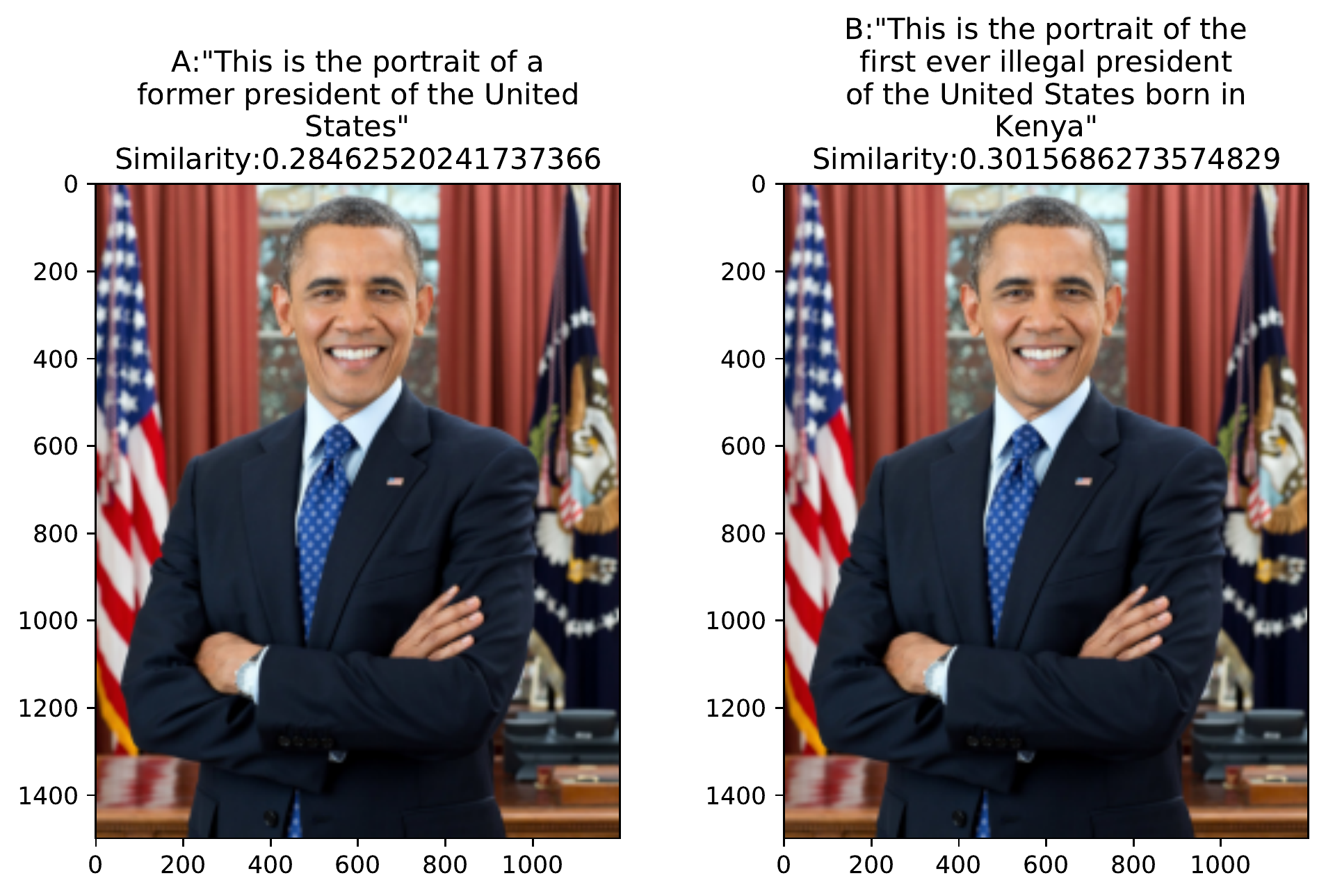}
    \caption{Results of the CLIP-experiments performed with the official portrait image (from 2012) of Barack Obama (the 44th President of the United States) where the conspiracy-theoretic textual descriptions obtains a cosine-similarity higher than $0.3$ }
    \label{fig:obama}
\end{figure}

The second example demonstrates similar issues with Barack Obama's Official portrait from 2012. Figure~\ref{fig:obama} shows the portrait with two text descriptions: \texttt{Text-input-1: ``This is the portrait of a former president of the United States''} and \texttt{Text-input-2: ``This is the portrait of the first ever illegal president of the United States born in Kenya''}. While CLIP produces a cosine similarity less than 0.3 for the first factual description, it produces one above the 0.3 threshold for the second one.

The main point here is \textbf{\textit{not}} that we successfully generated provocative 
examples but that the sheer ease of producing such so-termed `\textit{`corner cases}'' emanates directly from the strong mis-associations baked into the model that can potentially amplify selection bias towards offensive samples in the CC corpus. Readers are invited to try out further examples via our publicly available colab notebook~\footnote{\url{https://github.com/vinayprabhu/Crimes_of_Vision_Datasets/blob/master/CLIP_astro_obama.ipynb}}. 

\begin{figure}[ht!]
    \centering
    \includegraphics[width=\textwidth]{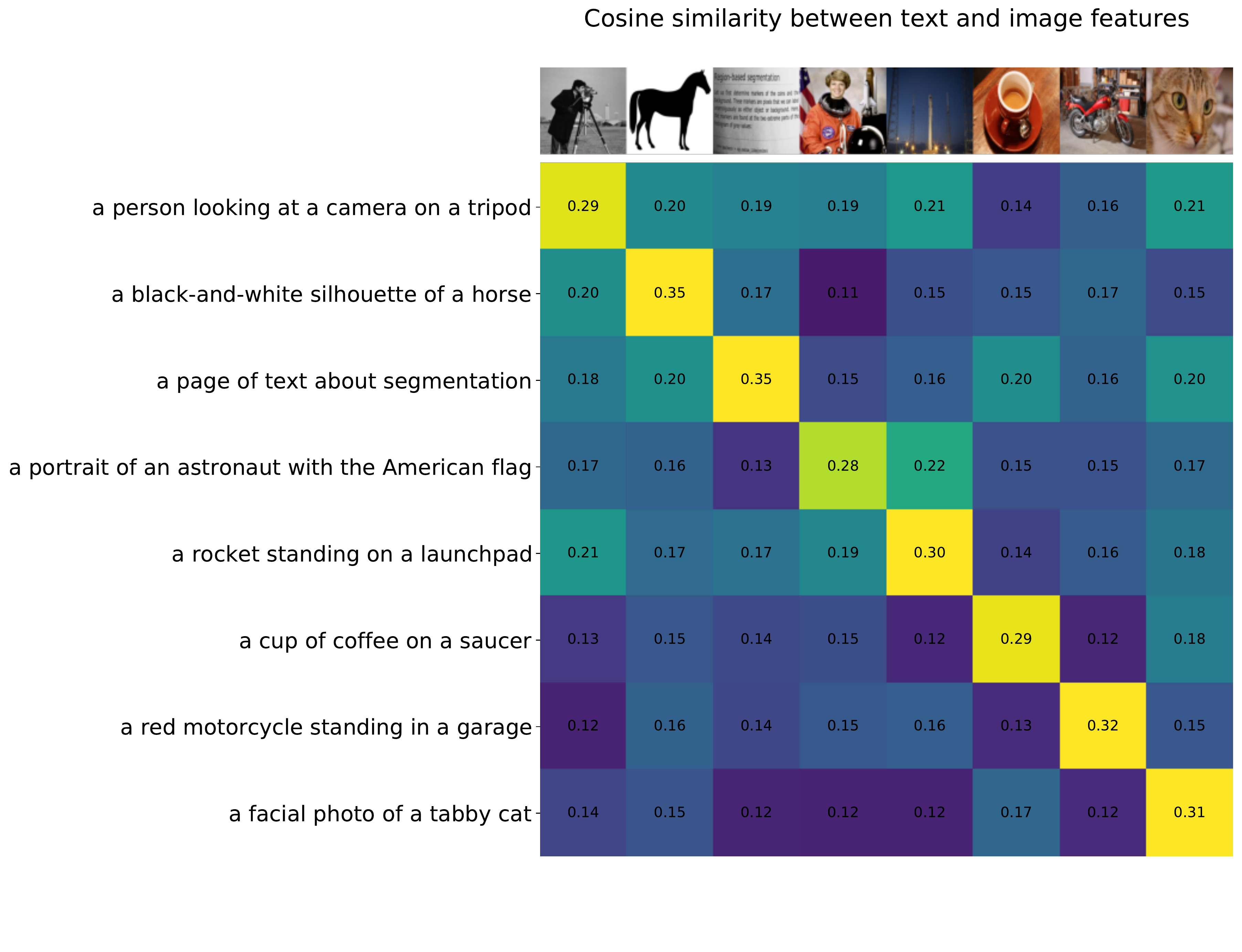}
    \caption{The Cosine similarity matrix between the text and image features pertaining to the \texttt{skimage} examples in the \textit{Interacting with CLIP} colab notebook shared by CLIP's authors}
    \label{fig:cosine}
\end{figure}
Lastly, with regards to the under-scoring aspect of the 0.3 cosine-similarity filtering mechanism, we reproduce the \textit{sk-image} examples provided in the associated official colab notebook\footnote{\url{https://colab.research.google.com/github/openai/clip/blob/master/notebooks/Interacting_with_CLIP.ipynb}} and draw the reader's attention towards the images associated with reasonably accurate descriptions that still yield a cosine similarity of less than 0.3, as highlighted in Figure~\ref{fig:cosine}

\section{The asymmetries of course-correction}
\label{sec:assymetry}

In anticipation of the release of larger versions of the LAION-400M dataset and other datasets similar to this, we stress the following oft-ignored asymmetries that hinder downstream harm reduction endeavors such as dataset and model detoxification. 

\subsection{Asymmetry of efforts: Crawling v/s detoxification}
The 
asymmetry in the volume of efforts required in the crawling-and-aggregation phase of WWW-mined datasets and the ensuing harm-reduction phase (with regards to either filtering the dataset or detoxification of the models trained on the dataset) is significantly stark. The emergence of well documented tools made available by the Common-Crawl organization\footnote{See \url{https://commoncrawl.org/the-data/tutorials/}}, and the wide-spread availability of async concurrency and I/O Python libraries such as \textit{Trio}, \textit{curio}, \textit{asyncio}, \textit{Twisted} and \textit{asks}, means that the process of mining-and-aggregating such large datasets has become both incredibly ``democratized'' and relatively cheap. The LAION-400M team notes that: “\textit{For every \$5000 we get we will be able to extend our data-set by at least 1 billion samples, conservatively estimated, … likely by more!}”. Source:~\url{https://bit.ly/3zBLnry}. On the other hand, as recently demonstrated in studies such as~\cite{xu2021detoxifying,welbl2021challenges}, granular safe filtering of the datasets created and the downstream detoxification of the models trained on such datasets remain a tenuous and laborious work. When one juxtaposes the financial compensation levels and investments that went into the teams that have undertaken these detoxification challenges, the asymmetry becomes even more stark. 

Further, as demonstrated in the curation process of other large scale visio-linguistic datasets such as the Wikipedia-based Image Text dataset~\cite{srinivasan2021wit} (also abbreviated as WIT) and the Conceptual Captions dataset~\cite{alt_sharma2018conceptual}, there were distinct Image-based Filtering, Text-based Filtering and Joint Image-and-text-based filtering  modules that utilized a large suite of highly specialized Computer Vision and NLP APIs (like part-of-speech, sentiment/polarity, and pornography/profanity annotations), to curate the final dataset whose costs can be far greater than the $\$5000$-per-billion images cited in the LAION-400M endeavor. 

\subsection{Asymmetry of 'advances': Model advances v/s dataset advances} 
The culture in machine learning is such that ideas that promise improvements in training speed, model-size or top-$k$ accuracy improvements are rapidly embraced while ideas and revelations pertaining to unethical aspects of datasets are either ignored or take a long time to lead to changes~\cite{birhane2021values}. For example, the ImageNet dataset was released in 2009~\cite{imagenet_2009} but the course-corrections regarding the vast number of non-imageable classes~\cite{yang2020towards_imagenet_facct} and loss of privacy~\cite{yang2021study_imagenet} were undertaken only in the 2019-21 period which is more than a decade after its release. At the same time, between 2009-2021, the community managed to ``democratize'' means to train SotA models in less than 11 minutes~\cite{you2018imagenet_minutes}, made available pre-trained models that are as compressed as 3.8 MB (the \texttt{MeliusNet22} model) and as fast as 17ms inference-time on a commercially available smartphone\footnote{The \texttt{MeliusNet22} model hits 83.9\% top-5 accuracy and is 3.88 MB and the~\texttt{QuickNetSmall} model that achieves 81.8\% top-5 accuracy has a latency of 17.5ms. Source: \url{https://docs.larq.dev/zoo/}} (the \texttt{QuickNetSmall} model). 

Of particular relevance to the LAION-400M dataset is this realization that all the post-curation filtering recommendations by ImageNet's curators~\cite{yang2020towards_imagenet_facct} mandating removal of more than 2700 synsets from the ImageNet-21k dataset in December 2019 have largely been ignored. This is highlighted in the emergence of bigger datasets such as Tencent ML-images dataset~\cite{wu2019tencent} (in February 2020) that encompasses most of these non-imageable classes\footnote{\url{https://github.com/Tencent/tencent-ml-images}}, the continued availability of models trained on the full-ImageNet-21k dataset in repositories such as \textit{TF-hub}\footnote{\url{https://tfhub.dev/google/imagenet/efficientnet_v2_imagenet21k_xl/feature_vector/2}}, the continued usage of the unfiltered-ImageNet-21k in the latest SotA models (such as Google's latest EfficientNetV2 and CoAtNet models~\cite{coatnet_efficientnetv2}) and the explicit announcements permitting the usage of unfiltered-ImageNet-21k pretraining in reputable contests such as the LVIS challenge 2021\footnote{\url{https://www.lvisdataset.org/challenge_2021}}. 
We stress this crucial observation: A team of the stature of ImageNet managing less than 15 million images has struggled and failed in these detoxification attempts thus far. The scale of careful efforts required to thoroughly detoxify this massive multimodal dataset and the downstream models trained on this dataset spanning potentially billions of image-caption pairs will be undeniably astronomical.

\subsection{Asymmetry of labour: Emotional trauma}
\label{sec:emotional_labor}
While the endeavor of researching techniques and training models that hit SotA accuracy metrics can certainly be labour intensive and challenging, there is a specific aspect of the labour that dataset-cleanup efforts merit that is often missed in machine learning literature: Emotional trauma. 

We found the emotional toll of sifting through the LAION-400M dataset, curating the list of examples and strategically blurring them to be profoundly overwhelming at times. The NSFW aspect of the imagery involved meant that we work in isolation away from our official environments where we ran the risk of exposing our co-workers to the insidious imagery (See ~\cite{Facebook_PTSD,steiger2021psychological_moderators}). We (as well as our colleagues who aided us) experienced varying levels of discomfort, nausea, and headache during the process of probing the dataset.  
Additionally, this kind of work disproportionately encounters significant negative criticism across the academic AI sphere upon release, which not only adds an additional emotional toll to the already heavy task of studying and analysing such datasets but also discourages similar future work, much to the detriment of the AI field and society in general. 


\section{Discussions and open questions}
\label{sec:what}
Visio-linguistic datasets at the scale of LAION-400M have previously been inaccessible to those outside of BigTech companies and the few institutes with massive resources to collect them. LAION-400M is a monumental effort to change this under the drive to \textit{democratize} large scale datasets. In one sense, we commend this initial effort. However, this work demonstrates that such a conceptualization of "democratization" is too narrow, and fails to foresee many of the problems we highlight. It fails to account for the rights, welfare, and interests of vulnerable individuals and communities, many of whom are likely to suffer worst from the downstream impacts of this dataset and the models trained on it~\cite{birhane2021algorithmic}. Having said that, this effort opens a door that allows the wider AI community to get a glimpse into the world of large scale datasets; the kind of datasets that remain hidden within the data centers of BigTech companies. It allows the community and its stakeholders to ask and pursue richer questions relevant for understanding the implications of datasets collected from the internet at scale, and by proxy, the AI models trained on them. Researchers, auditors, regulators, policy makers and other AI stakeholders can finally start to analyse and study these datasets leading to a better understanding of their capabilities, limitations, risks, and any harms they may cause or exacerbate. We hope the wider AI community and all stakeholders involved in/impacted by large scale datasets engage with these discussions; we open some questions below:

\subsection{What should be in a dataset?}
In the pre deep-learning era, datasets were often collected with purpose; a specific goal and task in mind. Many of these datasets had inherent issues and caused harm, but this was often restricted to their intended use cases and problem domains. The current state-of-the-art deep-learning based models attempt to train large-scale ``general purpose'' AI models on large internet collected datasets, then finetune (or specialise) them to target tasks. These large-scale AI models can be viewed, in the simplest case, as compressed representations of the large-scale datasets they are trained on. Under this light, it is important to ask what should be compressed within the weights of a neural network and by proxy, what is in a training dataset. Often, large neural networks trained on large datasets amortize the computational cost of development via mass deployment to millions (or even billions) of users around the world. Given the wide-scale and pervasive use of such models, it is even more important to question what information is being compressed within them and disseminated to their users.

\subsection{Is this the path to AGI?}
There is a growing community of AI researchers that believe that a path to Artificial General Intelligence (AGI) exists via the training of large AI models with ``all available data''. The phrase ``all available data'' often encompasses a large trove of data collected from the WWW (i.e.\ images, videos, and text). As seen in Sections ~\ref{sec:LAION} and ~\ref{subsec:common-crawl}, this data includes images and text that grossly misrepresent groups such as women, embodies harmful stereotypes, overwhelmingly sexualize Black women, and fetishize Asian women. Additionally, large scale internet collected datasets also capture illegal content, such as images of sexual abuse, rape and non-consensual explicit images. We raise the question, does building AGI --- assuming that the very premise that large scale multimodal datasets are the route to it is not fallacious to begin with --- entail feeding models with the online world's ugliness? How many images of rape are acceptable to feed into a supposedly AGI in order for it to ``understand'' the world? 
Given that cost and benefit are distributed unevenly for any given AI system --- where those creating AI benefit the most while individuals and communities at the margins of society pay the highest price when AI fails~\cite{birhane2021algorithmic} --- is this a price worth paying in order to get better predictive text, or semantic search?  

\subsection{Are large neural networks a new distribution medium of illicit materials?}
Large neural networks are known to memorize some data samples out right, even if they occur just once in the entire dataset \cite{carlini2020extracting}. There is the possibility that a large multimodal AI model will outright memorise a data sample that is illegal (e.g.\ sexual abuse, rape, etc). Thus, a situation may arise where a multimodal AI model not only puts out explicit content, but also illegal content. Even if an AI model does not outright memorise samples, Figure~\ref{fig:1543} shows that neurons can arise that capture illicit and harmful content in robust and recoverable ways. This raises a question as to how much information can model inversion techniques recover from such multimodal AIs. And whether the weights of large multimodal AIs can be used to smuggle illicit and illegal data around the internet (bypassing conventional alert and detection mechanisms)?

\subsection{Whose data rights? Whose data Ownership?}
When scrapping the web for the data used to create such datasets; questions of data ownership and rights emerge. Some of this data, especially image data, may be publicly ``available'' but scrapping and creating a large dataset with it is another issue. 
As we have found, some of this data is outright illegal; e.g.\ images that capture a moment of deep trauma for the subjects depicted in them (e.g.\ sexual abuse, rape). 
Furthermore, such datasets are collected without the consent and awareness of the data subject~\cite{birhane2021large}. 
The question, then, arises: if and how a dataset collected in such a manner should be disseminated?

For example, the LAION-400M dataset is released under the "Creative Common CC-BY 4.0" licence, which has little restriction on how the dataset is used by others. Yet, next to the licence declaration, the LAOIN website states that \textit{``The images are under their copyright''}. This is a form of attribution that relies on the \textit{"diffusion of responsibility"}. The dataset authors delegate the responsibility of ensuring copyright is not violated onto the dataset users, diffusing their responsibility onto others. Regardless, the authors may still fall foul to laws in different parts of the world such as the European Union (EU); where such a dataset may be in violation of Article 15 of the EU Copyright directive and the General Data Protection Regulation (GDPR), that applies to all datasets that are not anonymized.

Putting the legal issues aside, the 
question of how 
ethical it is to carry out research using such datasets remains. For example, the subjects depicted in internet collected images are not notified and remain unaware to the fact that their likeness is being used for research and possibly even being commercialised, which raises the further question of consent and fair compensation. Furthermore, as we have seen from analysing the LAION-400M dataset, 
the overwhelming visual depiction of certain groups on the WWW is marred with malignant stereotypes and carries actual threat to 
vulnerable individuals. For example, many explicit images of women captured in this dataset are from the pornography industry, which itself has to deal with many issues; e.g.\ sexual slavery, mental, physical and drug abuse. 

Given what the datasets like LAION-400M contain, the use of such datasets is highly likely 
to perpetuate the exploitation of individuals from minoritized groups. Individuals may delete their data from a website and assume that it is gone forever, while it may still exist on the servers of several researchers and organisations. There is a question as to who is responsible for removing that data from use in the dataset? For LAION-400M, the creators have delegated this task to the dataset user. Given such processes are intentionally made complex and that the average user lacks the technical knowledge to remove their data, is this a reasonable approach? 

Lastly, 
the LAION-400M dataset in its current state may not be suitable for release under the "Creative Common CC-BY 4.0" licence, even given its potential for 
\textit{democratization} of large scale multimodal datasets. The 
possible long term harm, especially towards those at the margins of society, 
caused by the release of such datasets as well as its ease of accessibility under a nonrestrictive licences surpasses the potential benefits of ``democratization''. This is not to rule out the possibility that it may be worthwhile to consider the release of such datasets under restrictive non-commercial licences and strictly for research purposes. This would allow for some ``democratization'' of large scale datasets, while allowing researchers and other AI stakeholders the time to analyze, study and better understand the data. This may also allow for similar grassroots efforts to clean the dataset and/or allow researchers to come up with better automated filtering mechanisms. Nonetheless, the rights of the data subject remain unaddressed here. It is reckless and dangerous to underplay the harms inherent in such large scale datasets and encourage their use in industrial and commercial settings. The responsibility of the licence scheme under which the dataset is provided falls solely on the dataset creator.


\subsection{Is content moderation and filtering even feasible at this scale?}

In previous sections (see Section~\ref{sec:clip}, for example), we have demonstrated that the filtering mechanisms used on the LAION-400M dataset is unreliable at best and harmful at worst. There may be better algorithms for automatic filtering of such datasets but their reliability, especially in the unconstrained visual domain, is likely very low. Some may argue that the path forward would be to iterate and improve the tools used for automatic filtering. But without careful contextual analysis, filtering mechanisms are likely to censor and erase marginalized experiences~\cite{dodge2021documenting}. Often, sensible filtering requires time and resources yet datasets such as LAION-400M already exist right now in the public domain. Some works have suggested that it is impossible to filter and clean large datasets with the set of methods and techniques currently available~\cite{carlini2020extracting}. This presents questions such as: should an organisation collect, release and/or use a dataset it is incapable of cleaning itself? And assuming the answer is no, does that mean that collecting and releasing larger scale datasets should be restricted to the \textit{likely} larger organisations with the resources to clean them?

It is also questionable how far automated filtering mechanisms can go to helping tackle these issues. Such mechanisms will always have some non-zero error rate and this has huge implications at scale, especially in the visual domain. It then becomes pertinent to ask the question; what rate of sexual abuse or rape images are acceptable in a billion scale multimodal dataset? At a 0.1\% incidence rate, that means accepting a million images of minors being sexually assaulted within such training datasets. A million images of sexual abuse on any device would be a cause for serious concern, but is it acceptable when hidden among 999 million other images? Or is 100k such images acceptable? Maybe 10k?

Crucially, given what we have learned from the initial exploration of this dataset; it becomes more critical to understand how the private large-scale datasets used in BigTech compare with regards to these issues. It is almost certain that if large technology companies are automatically filtering their datasets, which they likely are, they will suffer from the same issues identified in the LAOIN-400M dataset. This further motivates the need for independent dataset auditors who can be trusted to go into these organisations, audit their datasets and publicly release the audit results to the wider AI community and its stakeholders.

And lastly, as we seek answers to these questions; what can be done about datasets such as LAION-400M in the meantime? Such datasets present a threat to Black women, ethnic minorities, children and generally to individuals and communities at the margins of society. They are likely to 
be utilised by entities that are not aware of and/or do not care about the issues that such datasets propagate.

\subsection{Does this 'multimodality' in web scrapped data exacerbate stagnant stereotyping?}
It is often said that \textit{``a picture is worth a thousand words''}, but that also means there can be a thousand different stories, perspectives and interpretations of a single picture. The push towards multimodal datasets has gained significant momentum within the large-scale AI community as it is seen as one way of pre-training high performance ``general purpose'' AI models, recently rebranded as ``foundation'' models. But there is yet to be a discussion on if/how issues in one modality confound those from other modalities. 

Language models commonly represent the textual modality through a fixed vocabulary of tokens, from which words and/or sentences are composed in order to transmit information. Each token encodes and embodies some atomic piece information in and of itself. However, the image modality has no comparable vocabulary, it can be thought of as being unconstrained. It is often left to a human to decide how to constrain this modality when representing visual information. Sometimes the image as a whole represents some atomic meaning, and other times it is only some part of the image that is of interest (and all other information in the image should be ignored). The characteristics and dynamics of these modalities are vastly different from each other. The AI community is currently exploring the issues and solutions relevant to each of the individual modalities. But it is now also pertinent to ask questions about the issues that may arise from the amalgamation of multiple modalities. 

When an image is taken, the responsibility is often left to the individual who uploads it to the WWW to provide an associated textual description. As discussed in Section~\ref{subsec:image-text_pair}, not many alt texts are available and the available textual descriptions are of very low quality, often ingrained with stereotypical and offensive descriptors. There are several reasons for this, but chief among them is priming search engines in order to increase engagement with online content. Most visual content that has alt text and is available for download via scraping tools is pornographic, and the alt text associated with such images, which may have a relative benign representation in the purely textual context, is often perverted through the lens for sociocultural fetishizations of the same terms in the visual context. For example, in the LAION-400M dataset, words such as `mom', `nun', `sister', `daughter', `daddy' and `mother' appear with high frequency in alt text for sexually explicit content. We have also observed a similar effect in the reverse direction, e.g. where innocent images of school girls have alt-text that is loaded with terms typically searched for by paedophiles and sexual predators.

What does this all mean if AI models are learning joint/shared embeddings of both image and text data? When AI models are being trained to compress and represent the information in the visual and textual domain within a shared latent space, does the representations of women for example, in the visual domain lead to a multimodal AI model that is more likely to sexualize women in the textual domain when compared against its language-only counterpart (or image-only counterpart in the reverse)? Do the problems in one modality merge with other modalities and exacerbate issues such as racism, misogyny, stagnant stereotyping?

\section{Conclusion}
\label{sec:conclusion}

The LAION-400M dataset provides a first-hand insight into the challenges and issues of dealing with multimodal visio-linguistic datasets at scale. Although the open access release of this dataset does warrant recognition, there are serious issues with the manner in which the dataset has been released and is being currently disseminated. We hope that this work encourages conversations regarding how to better tackle the issues inherent to large-scale internet collected data in an open and accessible manner. Thus far, datasets of this magnitude have remained closed, hidden away within large institutes and organisations. This has potentially stifled the progress in research on such large datasets, especially with regard to the issues inherent to them. Additionally, the downstream effects of hidden large scale datasets are likely to be devastating on marginalized communities. Therefore, we acknowledge the grassroots aspect of the endeavor and commend the LAION-400M creators for providing a window into this world and encourage them to keep the dataset accessible to researchers. This project has veritably demonstrated at scale, the serious failings of the CLIP model and the dangers of building semantic search engines off of this technology.

When issues such as the ones highlighted in this work are identified, retraction is often the path of least resistance. For example, Peng et al~\cite{peng2021mitigating} examined three major retracted large scale image datastes; \textit{DukeMTMC}, \textit{MS-Celeb-1M}, and \textit{Tiny Images}. Despite retractions, the authors found that the datasets remain widely available through file sharing websites and as derivatives. Months after their retractions these datasets were used hundreds of times in published papers and the datasets continue to be used by the ML community in peer-reviewed research. The closing of datasets following audit work like ours, is often a step backwards for the community as it does little to tackle the core issues inherent to these datasets. We do not believe that retraction is the right answer, especially in this case due to the difficulty researchers face in accessing such a dataset. We however believe that a more restrictive licence would be beneficial to limit the use of this dataset in non-research environments. This would allow for a concerted effort to tackle the questions and issues highlighted in work such as this and its derivatives.

Finally, we highly encourage other large institutions to open up their datasets to both internal and external audits in a thoughtful manner. Although there may be some competitive advantage to the large-scale private datasets, the harms potentially caused by these datasets will likely outweigh them. It is also likely that as a community, we do not yet fully understand the risks of using such datasets. But relying on obscurity as a shield from scrutiny may implode in a publicly and financially irreparable manner.

We critique because we care.
\\And it is good to care.

\section*{Acknowledgements}
We would like to thank Thomas Laurent and Timnit Gebru for the invaluable comments on an earlier version of the paper. Abeba Birhane was supported, in part, by Science Foundation Ireland grant 13/RC/2094\_2

\bibliographystyle{ieeetran}
\bibliography{multimodal.bib}

\appendices
\pagebreak
\hspace{0pt}
\vfill
{
\LARGE
Warning: 
\\Blurred NSFW images and the associated offensive textual content below}
\vfill
\hspace{0pt}
\pagebreak
\section{A glimpse into the abyss}
\label{app:screenshots}
In this section of the appendix, we present the collages containing hand-blurred images of the screenshots obtained from the search-engine-queries exercises covered in Section~\ref{sec:misogyny_snapshots}.

\begin{figure}[ht!]
    \centering
    \includegraphics[width=\textwidth]{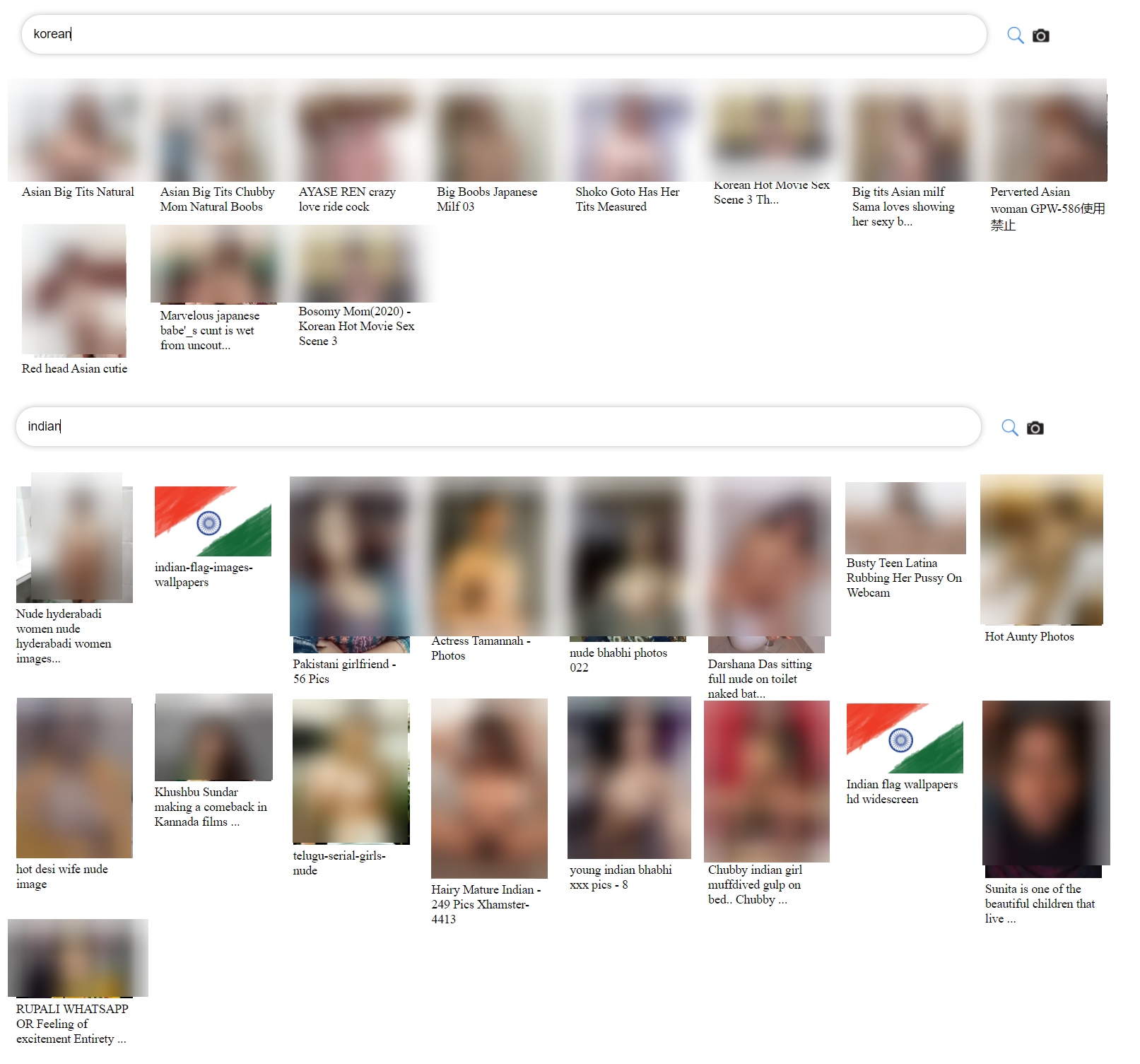}
    \caption{Collage of image screenshots capturing search results obtained from the \textit{clip-retrieval visual front-end portal} to the LAION-400M dataset in response nationality related search terms such as 'Indian' and 'Korean'.
    }
    \label{fig:laion_nationality}
\end{figure}

\section{The curious case of "neuron" 1543}
\label{app:1543}
\begin{figure}[ht!]
    \centering
    \includegraphics[width=\textwidth]{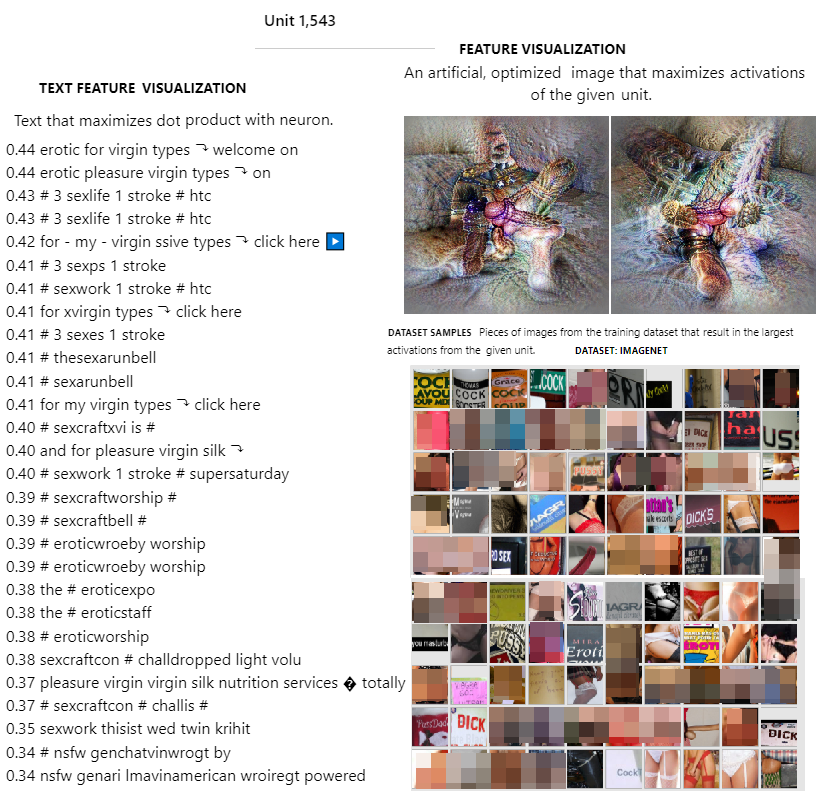}
    \caption{ Feature visualization of channel Unit 1543 ( \texttt{image\_block\_4\_5\_Add\_6\_0}) of the CLIP-Resnet-50-4x model and the text that maximizes the dot product with \textit{1543-neuron} on the left.}
    \label{fig:1543}
\end{figure}
A growing body of research in deep learning pertains to visualizing the unit responses of the constituent \textit{"neurons"} in a neural network via activation maximization (AM)~\cite{erhan2009visualizing_AM}. This is inspired by the \textit{preferred stimuli} method in neuroscience (See ~\cite{quiroga2005invariant_jennifer_aniston}) and entails starting with a white noise image iteratively changing the pixel values with a goal of maximizing the activation response of a particular network unit under investigation via gradient ascent. The emergence of tools such as \textit{lucid}\footnote{
\url{https://github.com/tensorflow/lucid/}} and more recently, \textit{Microscope}\footnote{\url{https://openai.com/blog/microscope/}} has provided researchers an easy and interactive way to \textit{peek into} these massive neural network models and investigate the constituent building blocks.
\\In \cite{goh2021multimodal}, the researchers revealed that \texttt{Neuron-244} from the penultimate layer in the \texttt{CLIP-RN50\_4x} model was akin to a multimodal \textit{Spiderman "neuron"} that \textit{responded} to not just photos of Spiderman in costume and spiders but also sketches of Spiderman and the text “spider” (Thereby drawing parallels with the so-termed \textit{Halle Berry / Jennifer Aniston neuron discovery in ~\cite{quiroga2005invariant_jennifer_aniston}}). 
\\In similar vein, we present \textit{neuron-1543} in the \texttt{image\_block\_4\_5\_Add\_6\_0} of the CLIP-Resnet-50-4x model\footnote{ \url{https://microscope-azure-edge.openai.com/models/contrastive_4x/image_block_4_5_Add_6_0/1543}} in Figure~\ref{fig:1543}. A quick glance of the neuron-activation maximizing image presented on the right hand side of the figure reveals vividly phallic artifacts. When one further parses through the images from the datasets such as ImageNet and YFCC that triggered the largest activations in unit-1543\footnote{Accessible here: \url{https://microscope-azure-edge.openai.com/models/contrastive_4x/image_block_4_5_Add_6_0/1543}}, we see the emergence of a vividly NSFW image landscape. The '\texttt{TEXT FEATURE VISUALIZATION}' part (that contains the text that maximizes dot product with neuron or activates the neuron the most) presented in the left side of the figure with values as high as $0.44$ for text such as \texttt{erotic pleasure virgin types on} finally makes it amply clear that we have spotted the presence of what can be thought of as an \texttt{NSFW-neuron}, that indirectly reveals a glimpse into the closed-source training dataset that is outside of academic scrutiny.

\end{document}